\documentclass{article} 
\usepackage{collas2025_conference,times}
\usepackage{easyReview}

\usepackage{amsmath,amsfonts,bm}

\def\eqref#1{equation~\ref{#1}}

\def\1{\bm{1}}

\DeclareMathAlphabet{\mathsfit}{\encodingdefault}{\sfdefault}{m}{sl}
\SetMathAlphabet{\mathsfit}{bold}{\encodingdefault}{\sfdefault}{bx}{n}

\DeclareMathOperator*{\argmax}{arg\,max}

\usepackage{hyperref}
\hypersetup{
    colorlinks=true,
    linkcolor=red,
    filecolor=magenta,
    urlcolor=blue,
    citecolor=purple,
    pdftitle={Overleaf Example},
    pdfpagemode=FullScreen,
    }
\usepackage{wrapfig}
\usepackage{subcaption}
\usepackage[capitalize]{cleveref}
\usepackage{algorithm}
\usepackage{algorithmic}
\usepackage{xcolor,colortbl}
\usepackage{multirow}
\usepackage{booktabs}
\usepackage{bbding}

\usepackage{pifont}
\usepackage{graphicx} 
\usepackage{caption}   
\definecolor{mygray}{gray}{0.4}
\definecolor{LightPurple}{rgb}{0.88,0.88,1}
\definecolor{blanchedalmond}{rgb}{1.0, 0.92, 0.8}
\definecolor{celadon}{rgb}{0.67, 0.88, 0.69}

\title{Query Drift Compensation: Enabling Compatibility in Continual Learning of Retrieval Embedding Models}

\author{\quad\quad\quad Dipam Goswami\textsuperscript{1,2} \quad Liying Wang\textsuperscript{1,2} \quad Bartłomiej Twardowski\textsuperscript{1,2,3} \quad Joost van de Weijer\textsuperscript{1,2}  \\ 
\quad\quad\quad \textsuperscript{1} Computer Vision Center, Barcelona, Spain  \\
\quad\quad\quad \textsuperscript{2}Department of Computer Science, Universitat Autònoma de Barcelona, Spain \\
\quad\quad\quad \textsuperscript{3}IDEAS Research Center, Warsaw, Poland \\  
{\quad\quad\quad \tt\small \{dgoswami, liying, btwardowski, joost\}@cvc.uab.es}
}

\collasfinalcopy 

\begin{document}

\maketitle

\begin{abstract}
Text embedding models enable semantic search, powering several NLP applications like Retrieval Augmented Generation by efficient information retrieval (IR). However, text embedding models are commonly studied in scenarios where the training data is static, thus limiting its applications to dynamic scenarios where new training data emerges over time. IR methods generally encode a huge corpus of documents to low-dimensional embeddings and store them in a database index. During retrieval, a semantic search over the corpus is performed and the document whose embedding is most similar to the query embedding is returned. 
When updating an embedding model with new training data, using the already indexed corpus is suboptimal due to the non-compatibility issue, since the model which was used to obtain the embeddings of the corpus has changed. While re-indexing of old corpus documents using the updated model enables compatibility, it requires much higher computation and time. Thus, it is critical to study how the already indexed corpus can still be effectively used without the need of re-indexing. In this work, we establish a continual learning benchmark with large-scale datasets and continually train dense retrieval embedding models on query-document pairs from new datasets in each task and observe forgetting on old tasks due to significant drift of embeddings. We employ embedding distillation on both query and document embeddings to maintain stability and propose a novel query drift compensation method during retrieval to project new model query embeddings to the old embedding space. This enables compatibility with previously indexed corpus embeddings extracted using the old model and thus reduces the forgetting. We show that the proposed method significantly improves performance without any re-indexing. Code is available at \url{https://github.com/dipamgoswami/QDC}.
\end{abstract}

\section{Introduction}
Information Retrieval (IR) is widely used in several NLP applications like semantic search and Retrieval-Augmented Generation (RAG) for LLMs. Text embeddings which encode a sentence or a chunk of text to low-dimensional embedding vectors are commonly used for these applications~\citep{lewis2020retrieval,ram2023context,izacard2023atlas}.
Semantic search enables us to retrieve most relevant responses from a document corpus for a given query. While non-semantic lexical approaches like TF-IDF and BM25~\citep{robertson2009probabilistic} were traditionally used for retrieval, dense retrievers like transformer architectures~\citep{vaswani2017u} are now widely used for semantic search~\citep{cer2018universal,yates2021pretrained,thakur2021beir,muennighoff2022sgpt,nussbaum2024nomic}. While lexical methods consider queries and documents as bag-of-words, dense retriever models encode the queries and corpus documents in a shared semantic embedding space~\citep{gillick2018end} which enables us to precompute and index the document embeddings from the corpus before performing retrieval. The standard practice~\citep{thakur2021beir,muennighoff2022sgpt,nussbaum2024nomic} is to consider a static setting in which the retriever model is trained on several datasets and the corpus documents are indexed using the trained retriever model. These indexed corpus document embeddings are then used for evaluation of retrieval for each task or dataset separately. However, in many practical applications not all datasets are jointly available, and new data arrives over time.
In order to improve the retrieval models on newly incoming datasets, one needs to continually train them. In this work, we study how the retriever models can be fine-tuned on new datasets over time and how updating the model can impact the retrieval process and performance.

Continual Learning (CL) enables neural networks to learn a sequence of tasks one after another and perform well on all seen tasks. Several research works~\citep{de2021continual,masana2020class,bornschein2023nevis,wang2024comprehensive,zhou2024class,verwimpcontinual} explore various aspects of CL, primarily for image classification tasks. A major challenge in CL is catastrophic forgetting~\citep{mccloskey1989catastrophic,kemker2018measuring} which refers to forgetting old tasks after learning new tasks. We focus on the task-incremental learning setting~\citep{van2019three}, where the task-id information is available during inference or retrieval. In this work, we discuss how continually updating the retriever models for document retrieval could lead to issues of non-compatibility between query and corpus document embeddings for old tasks. This is due to the fact that during retrieval of queries from old tasks, the query embeddings are obtained using the updated model while the corpus embeddings were previously indexed using the old model from the respective tasks. The issue of non-compatibility has previously been studied in CL for image retrieval works~\citep{ramanujan2022forward,wan2022continual,biondi2023cl2r}. The drop in performance of old tasks when naively fine-tuning on new tasks can be attributed to the \textit{embedding drift} (also referred to as semantic drift~\citep{Yu_2020_CVPR}) between old and new embedding spaces, as we demonstrate in our experiments.

\begin{figure*}
    \centering
    \includegraphics[width=\textwidth]{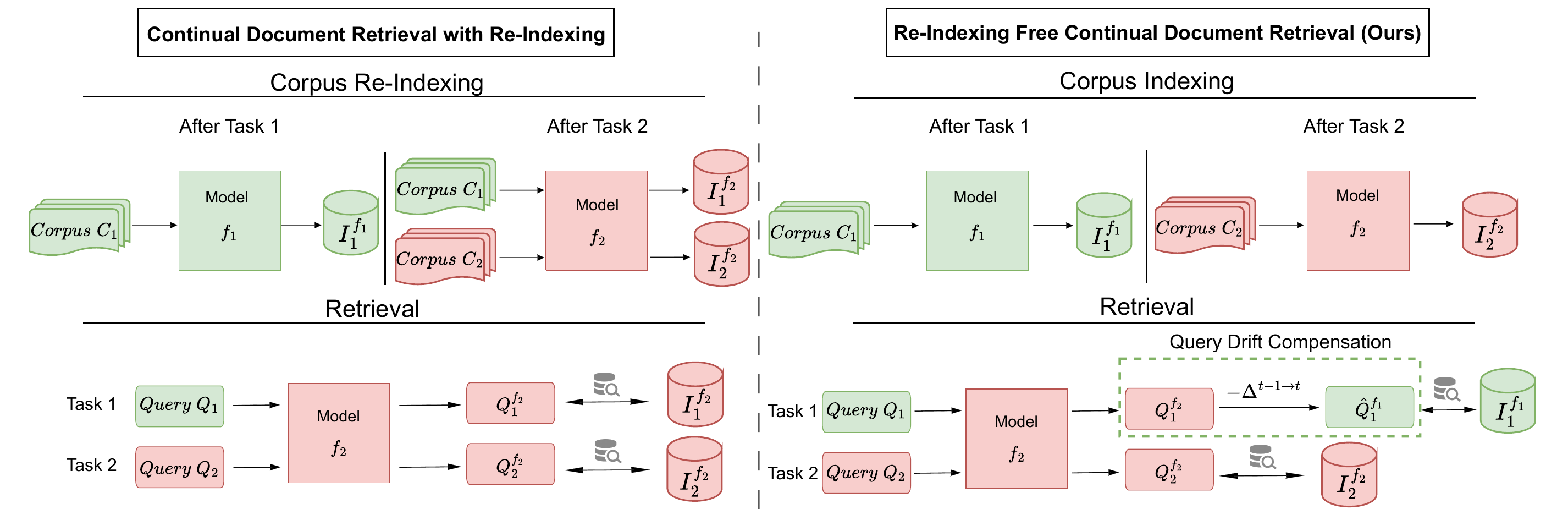}
    \vspace{-20pt}
    \caption{Continual Document Retrieval (CDR). Here, we consider a two-task continual setting and illustrate two different approaches to tackle the \textit{embedding drift} issue for old task retrieval in CDR which arises due to non-compatibility between query embeddings from the updated model $Q_1^{f_2}$ (in red) and corpus embeddings indexed using the old model $I_1^{f_1}$ (in green). (left) A naive approach to make the query and corpus embeddings compatible is by re-indexing the corpus documents using the updated model. However, re-indexing large amounts of documents from all old tasks after updating the model on every new task is time-consuming and computationally expensive. (right) To avoid re-indexing, we propose to estimate embedding drift from old to new model $\Delta^{t-1\rightarrow t}$ and compensate the drift from $Q_1^{f_2}$ during retrieval. The proposed query drift compensation approach enables compatibility by projecting the query embedding back to the embedding space of $f_1$ without any need for expensive re-indexing.}
    \label{fig:CDR}
    \vspace{-10pt}
\end{figure*}

One naive approach could be re-indexing of corpus documents from all old tasks (also known as back-filling in image retrieval~\citep{shen2020towards}) to ensure that the query and the corpus document embeddings are in the same embedding space as shown in~\cref{fig:CDR} (left). However, re-indexing of all old corpus documents after each task would involve very high computation costs. 
In order to avoid the re-indexing of old documents, we propose a novel query drift compensation approach which projects the query embeddings from the new model back to the embedding space of their respective tasks as illustrated in~\cref{fig:CDR} (right). We estimate query drift vectors for each task transition using the query samples from training data of the current task which 
we use to approximate the query drift of previous tasks.
We compensate or subtract these drift vectors from new model query embeddings to move them back to the old space. Thus, instead of re-indexing old documents, we propose a simple query projection which makes the query and document embeddings compatible and thus reduces the forgetting of old tasks significantly. 

We propose a CL benchmark for document retrieval tasks where we train on new datasets in each task using query-document pairs (see~\cref{table:examples}) and aim to improve the document retrieval performance on all old and new tasks. We fine-tune the retriever model on several datasets from the BEIR benchmark~\citep{thakur2021beir} to improve retrieval on specific data or domains. 
We also propose to employ embedding distillation between new and the previous task model to reduce the \textit{embedding drift} and the forgetting of old tasks. Our contributions can be summarized as:
\begin{itemize}
    \item We establish a benchmark for Continual Document Retrieval (CDR) to evaluate the effect of continual training on large-scale datasets and observe forgetting of old tasks after fine-tuning on new tasks. We show that knowledge distillation using both query and corpus embeddings during fine-tuning can reduce the forgetting.
    \item We discuss how non-compatibility due to the \textit{embedding drift} hurts retrieval performance. We propose a novel approach - Query Drift Compensation (QDC) which estimates the embedding drift between tasks after training on a new task. During retrieval, we propose to compensate the drift from queries (extracted from the updated model) to enable compatibility with the document embeddings (indexed using the old model).
    \item We demonstrate in our experiments how the proposed QDC enables continual learning of retrieval models on new datasets without any re-indexing of old documents and performs similar to joint training. We also show that our approach outperforms re-indexing based CDR.
\end{itemize}

\section{Related Works}
\vspace{-5pt}
\noindent\textbf{Continual Learning.}
CL methods~\citep{de2021continual,masana2020class,wang2024comprehensive,zhou2024class} focus primarily on reducing catastrophic forgetting of old classes after learning new classes. CL methods can be classified into class-incremental, task-incremental and domain-incremental settings~\citep{van2019three}. In this work, we focus on the task-incremental setting where we have access to task-id during inference, since it is a more realistic setting for retrieval. Existing CL approaches can be broadly divided into replay-based, regularization-based, parameter isolation-based and prototype-based methods. Replay-based methods~\citep{Rebuffi_iCARL_2017_CVPR,belouadah2019il2m} store exemplars from old tasks and use them during training on new tasks. However, storing exemplars has several limitations due to data regulations and privacy issues, as discussed in~\cite{Goswami_ADC_2024_CVPR}. Regularization-based methods prevent updates of important weights for old classes~\citep{kirkpatrick2017overcoming} or
employ knowledge distillation approaches~\citep{li_LWF_2018_PAMI,douillard2020podnet} between previous and current model to preserve knowledge from previous tasks. Some methods~\citep{mallya2018packnet,serra2018overcoming} divide the model to learn task-specific parameters. Prototype-based methods~\citep{Yu_2020_CVPR,goswami2023fecam} store a prototype representation for all seen classes and classify samples based on distances to the class prototypes. Semantic drift compensation~\citep{Yu_2020_CVPR,Goswami_ADC_2024_CVPR,Alex_LDC_2024_ECCV} has been used to improve prototype-based methods by updating old prototypes. We use the concept of drift compensation for backward projection of queries from latest model to old model embedding space, thus enabling query-corpus compatibility.

\noindent\textbf{Document Retrieval.} 
Following success of large language models, the use of neural retrieval models~\citep{karpukhin2020dense,liang2020embedding,khattab2020colbert,luan2021sparse,muennighoff2022sgpt} has become more common than traditional lexical approaches~\citep{robertson2009probabilistic}, which have the lexical gap~\citep{berger2000bridging}. Dense retrieval models~\citep{gillick2018end} map queries and documents in a shared dense embedding space, and scores their relevance based on cosine similarity between query and document embeddings. Recent works~\citep{gunther2023jina,nussbaum2024nomic} use modified BERT~\citep{devlin2019bert} and perform MLM pretraining followed by unsupersived contrastive finetuning on large corpus of data and finally finetune on labelled datasets. In this work, we use the nomic embedding model from~\cite{nussbaum2024nomic} which outperforms several competitive models~\citep{izacard2021unsupervised,wang2022text,li2023towards,gunther2023jina} in retrieval tasks.

\begin{table}
\begin{center}
\caption{Examples of query-document pairs of different datasets from the BEIR~\citep{thakur2021beir} benchmark.}
\resizebox{\textwidth}{!}{
\begin{tabular}{c|c|c}
 Dataset & Query & Document \\
\hline
\rule{0pt}{2ex} 
  MS MARCO~\citep{bajaj2016ms} &  how much magnesium  & Kidney Beans. A cup of kidney beans contains 70 mg of magnesium and \\
 & in kidney beans &   is a great source of protein and fiber. More: How to Bake With Beans.  \\
 \midrule[0.05pt]
  NQ~\citep{kwiatkowski2019natural}  & what is the meaning of a crown & Crowns are the main symbols of royal authority.[21]  \\
\midrule[0.05pt]
  HotpotQA~\citep{yang2018hotpotqa}  & What compound, known as aqua fortis or & Nitric acid (HNO), also known as aqua fortis and  \\
  & spirit of niter is used in rocket propellant? & spirit of niter, is a highly corrosive mineral acid. \\ \midrule[0.05pt]
  FEVER~\citep{thorne2018fever}   & what team won the az state peach bowl & The 1970 Peach Bowl was a college football bowl game between   \\  
  &  & the Arizona State Sun Devils and the North Carolina Tar Heels . \\ \midrule[0.05pt]
 FiQA-2018~\citep{maia201818}   & How can a 'saver' maintain or increase & Personally, I invest in mutual funds. Quite a bit in  \\ 
 & wealth in low interest rate economy? & index funds, some in capital growth \& international. \\
\hline 
\end{tabular} 
}
\vspace{-10pt}
\label{table:examples}
\end{center}
\end{table}

\noindent\textbf{Continual Document Retrieval.}
In CDR, the retriever model is expected to continually learn from new tasks over time without forgetting the previous tasks.
While several works studied continual image retrieval~\citep{wan2022continual,biondi2024stationary} and continual multimodal retrieval~\citep{wang2021continual}, fewer studies have explored continual learning in information retrieval. \cite{Lov_CSinNR_2021_AIR} investigated the catastrophic forgetting problem in neural ranking models for information retrieval.
\cite{GeraldSoulier_CLforLTSinNIR_2022_AIR} built a continual information retrieval setting using a single dataset MS MARCO~\citep{bajaj2016ms} and observed that catastrophic forgetting exists in IR to a lesser extent than image classification tasks. 
\cite{cai_L2R_2023_ACMIKM} proposed a re-indexing free memory-based method for first-stage retrieval in a different setting with unlabeled new task documents. Recently,~\cite{hou2025advancing} investigated how existing CL methods work in CDR by dividing MS MARCO into multiple tasks based on topics. Another set of works~\citep{kishore2023incdsi,mehta2023dsi++} explored differentiable search index for CDR on how to encode the corpus of documents in the model parameters where the model output for a given query is the predicted document. In this work, we propose a more comprehensive benchmark for CDR  using five large-scale datasets from~\cite{thakur2021beir} and analyze how continual training affect the retrieval performance.

\section{Continual Learning for Document Retrieval}
Here, we introduce and formalize the setting of Continual Document Retrieval (CDR). Following~\cite{thakur2021beir}, we refer to a text of any length from the corpus as a `document'. Document retrieval aims to return the most relevant document $d$ from the given corpus $C$ as a response to the query $q$ provided by the user. In the continual setting, we refer to the set of queries and corpus of documents for task $t \in [1,T]$ as $Q_t$ and $C_t$. Here, we denote a single query and document from task $t$ as $q_t$ and $d_t$ respectively, where $d_t \in C_t$. In the first task, we train the embedding model $f_1$ on query-document pairs $\{Q_1, D_1\}$ from the training set of task $1$. Note that $D_1 \in C_1$ refers to the set of documents from the corpus of task 1 that are used for training (typically the entire corpus is larger). We use the trained model $f_1$ for indexing all the documents from $C_1$ and refer to the indexed document embeddings as $I_1^{f_1}$. Similarly, we also use $q^{f_i}_j=f_i(q_j)$ . Here, the superscript term denotes the model used for indexing and the subscript term denotes the task to which the document belongs. Similarly, for task $t$, we train embedding model $f_t$ on $\{Q_t, D_t\}$ pairs from the training set of task $t$ and the corpus documents indexed by $f_t$ are referred to as $I_t^{f_t}$. In our setting, we consider that during the training of task $t$ we have no access to any data from previous tasks (also known as exemplar-free continual learning~\citep{smith2023closer,petit2023fetril,goswami2023fecam,magistrielastic}).

\noindent\textbf{Non-compatibility due to Embedding Drift.} After task $t$, we have access to the updated model $f_t$ only, and thus we need to use $f_t$ to embed the queries from all tasks during retrieval phase. For an old task $t'<t$, the queries $Q_{t'}$ from task $t'$ are embedded using $f_t$ denoted as $Q_{t'}^{f_t}$ but all the corpus documents $C_{t'}$ from $t'$ were already indexed after task $t'$ using $f_{t'}$ and stored as $I_{t'}^{f_{t'}}$. We refer to this phenomenon as \emph{embedding drift}. This leads to non-compatibility between query and document embeddings, since they are in two different embedding spaces (the embedding space has been updated during the continual training). We illustrate the setting of continual document retrieval and the non-compatibility issue in~\cref{fig:CDR}.

\noindent\textbf{Re-indexing of old task corpus.} 
A naive approach to enable compatibility of query and document embeddings is to re-index all the corpus documents from all old tasks to obtain $I_{t'}^{f_t} \; \forall \; t'<t $ after training on a new task $t$. Re-indexing removes the embedding drift between the queries $Q_{t'}^{f_t}$ and corpus documents $I_{t'}^{f_t}$, which are then in the same embedding space thus resolving the non-compatibility issue (see~\cref{fig:CDR}).

However, re-indexing involves high computation costs and it potentially takes a considerable amount of time to re-index large amounts of documents from all old tasks after
finetuning on every new task. Therefore, in this paper, we focus on re-indexing free CDR. Re-indexing (or back-filling) based approach has been found to be effective and considered an upper bound in fine-tuning of image retrieval models~\citep{shen2020towards}.
Interestingly, from our analysis in CDR, we show that re-indexing based CDR performs poorly.
We attribute this to misalignment due to the unequal drift of query and corpus embeddings (similar to observations in multi-modal continual retrieval~\citep{wang2021continual}). We revisit and analyze this in the experiments section.

\section{Re-indexing Free Continual Document Retrieval}

\subsection{Training Strategy}
We follow the training strategy from~\cite{nussbaum2024nomic} which performs masked language modeling (MLM) pre-training to train a long-context BERT model followed by multi-stage contrastive learning~\citep{li2023towards}. The first stage is unsupervised contrastive pre-training using the InfoNCE contrastive loss~\citep{oord2018representation} on huge amounts of publicly available text-pairs which are mined from the web. Following the pre-training stages, the final step is performing supervised contrastive finetuning on each dataset at each task. Here, we consider the model pre-trained using MLM and unsupervised contrastive learning and then continually finetune the pre-trained model with supervised contrastive learning on query-document $(q,d)$ training pairs of each task. 

For the continual training at task $t$, we initialize a new model $f_t$ with the weights of $f_{t-1}$ and then perform contrastive training on $(q,d)$ pairs from the training set of task $t$. For each $(q,d)$ pair, we consider the most similar documents as hard negatives and use $\mathcal{H}$ hard negatives for each query which are mined from the same dataset. 
We minimize the contrastive loss function for a given batch of $(q,d)$ pairs as follows:
\begin{equation}\label{eq:1}
\mathcal{L}_{C} = -\frac{1}{n} \sum_{i} \log \frac{e^{S(f_t(q_i), f_t(d_i)) / \tau}}{\sum_{j=1}^{n} e^{S(f_t(q_i), f_t(d_j)) / \tau} + \sum_{h = 1}^{\mathcal{H}} e^{S(f_t(q_i), f_t(d_h)) / \tau}} ,
\end{equation}
where $S(q, d)$ is the cosine similarity, $\tau$ is the temperature parameter and $n$ is the batch size. Here, the first term in the denominator $\sum_{j=1}^{n} e^{S(f_t(q_i), f_t(d_j)) / \tau}$ includes the in-batch negatives when $j \neq i$ and the second term $\sum_{h = 1}^{\mathcal{H}} e^{S(f_t(q_i), f_t(d_h)) / \tau}$ includes the hard-negatives for the corresponding query.

\begin{figure*}
    \centering
    \includegraphics[width=\textwidth]{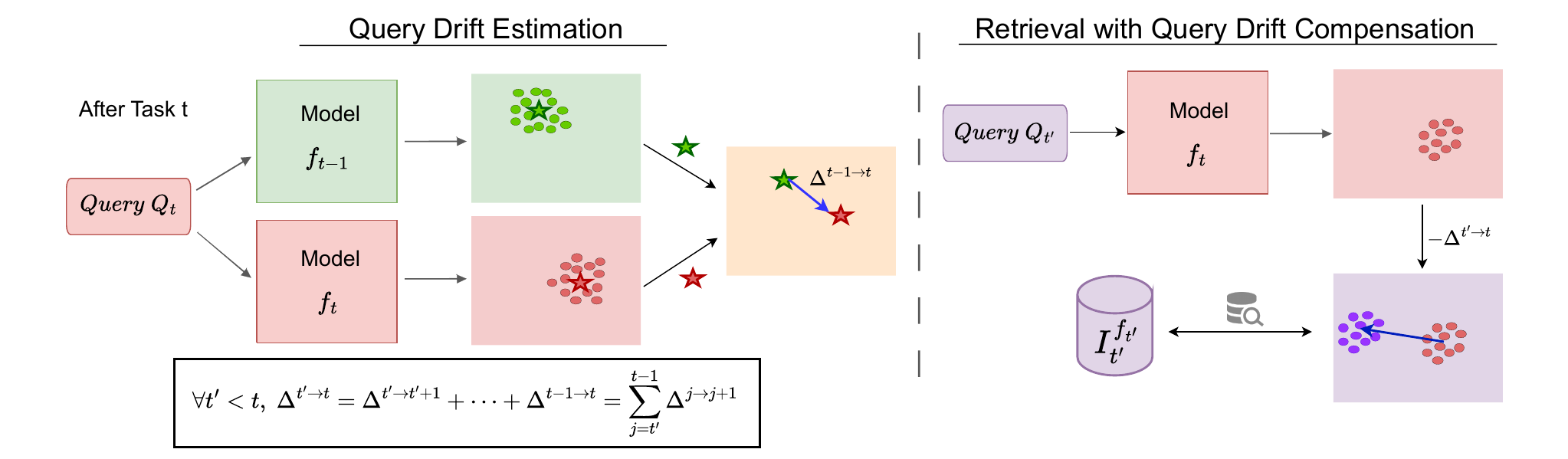}
    \caption{Illustration of Query Drift Compensation. (left) We show how to estimate the query drift vectors $\Delta^{t-1\rightarrow t}$ for each task transition (from $t-1$ to $t$) after training on task $t$. We store the drift vectors of all old tasks. For an old task $t'$ ($t' < t$), we obtain the drift vector $\Delta^{t'\rightarrow t}$ by addition of all drift vectors from task $t'$ to $t$. (right) During retrieval of old task $t'$, we compensate the query embeddings with drift vector $\Delta^{t'\rightarrow t}$ to project them from embedding space of task $t$ to that of $t'$. As a result, we compare the query and previously indexed document embeddings in the same embedding space of task $t'$ (in purple), thus avoiding the non-compatibility issue.}
    \label{fig:QDC}
    \vspace{-10pt}
\end{figure*}

\subsection{Embedding Knowledge Distillation}
A naive application of~\cref{eq:1} for all the tasks would lead to catastrophic forgetting and an embedding which is especially tailored to the last task. Therefore, here we adapt a commonly used regularization technique in CL for CDR to prevent forgetting~\citep{li_LWF_2018_PAMI}. More particularly, to reduce the feature drift and improve stability, we propose to perform feature distillation~\citep{Yu_2020_CVPR} by minimizing the cosine distance between the embeddings from old and new models. We perform the distillation on embeddings of both queries and documents from training data of the new task.
We minimize the following distillation loss for a given batch of $(q,d)$ pairs:
\begin{equation}
    \mathcal{L}_{D} = \frac{1}{n}\sum_{i=1}^n (D_c(f_t(q_i), f_{t-1}(q_i)) + D_c(f_t(d_i),f_{t-1}(d_i))) ,
\end{equation}
where $D_c$ is the cosine distance between embeddings and $n$ is the batch size.
Even though the distillation improves the stability of the network, it does not completely solve the non-compatability issue discussed before, since the learning of new tasks still requires the embedding space to adapt and incorporate new knowledge.

\subsection{Query Drift Compensation}
Here we propose Query Drift Compensation (QDC) which addresses the non-compatibility issue in re-indexing free CDR. In this case, for $t'<t$ we would like to query the corpus $I^{f_{t'}}_{t'}$ with the query embedding from the same embedding model $Q^{f_{t'}}_{t'}$, however we have access to the query embedding using the latest embedding model $Q^{f_t}_{t'}$, which leads to the non-compatibilty issue. 
To enable backward compatibility of old task query embeddings $Q_{t'}^{f_t}, \; \forall \; t'<t$ (indexed using continually updated new model $f_t$) with the corpus embeddings indexed using the old model $f_{t'}$, we propose to use drift compensation inspired by SDC~\citep{Yu_2020_CVPR} which was proposed for image classification in continual learning.

We define the difference between the query embeddings $q^{f_{t'}}_{t'}$ and $q^{f_{t}}_{t'}$ as the \emph{query drift}:
\begin{equation}
\delta^{t'\rightarrow t}=q^{f_{t}}_{t'}-q^{f_{t'}}_{t'} ,
\end{equation}
If we have this query drift, the desired embedding $q^{f_{t'}}_{t'}$ can be easily computed with  $q^{f_{t'}}_{t'}=q^{f_{t}}_{t'}-\delta^{t'\rightarrow t}$. However, we do not have access to $\delta^{t'\rightarrow t}$ as we cannot access the old task training data (the old task queries $q_{t'}$ cannot be used during task $t$). Therefore, in the following, we propose a method to approximate this drift based on the current task query drift.

\begin{algorithm}[tb]
\caption{Proposed Method for Continual Document Retrieval}
\label{alg:algo}
\begin{small} 
\begin{minipage}[t]{0.58\textwidth}
\begin{algorithmic}  
\STATE  \textbf{Continual Training}:
\STATE \textbf{Input:} 
\STATE $t \in [1,T]$: task number; $(Q_t, D_t)$: training data; $C_t$: corpus
\STATE $f_0$: pre-trained model; $\mathcal{H}, \tau$: hyper-parameters
\STATE \textbf{Output:} 
\STATE $f_t$: Model trained in task $t$
\STATE $\{I_z^{f_z}\} \; \forall \; z \in [1,t]$: Indexed corpus embeddings
\STATE $\{\Delta^{z\rightarrow z+1}\} \; \forall \; z \in [1, t-1]$: drift vectors
\vspace{2mm}
\FOR{task $t \in [1,T]$}
\STATE $f_t = f_{t-1}$    \quad\quad \textcolor{gray}{\# $f_{t-1}$ is frozen and $f_t$ is trainable}
\FOR{each $(q,d)$ pair in $(Q_t, D_t)$}
    \STATE $\mathcal{L}_{C} = -\frac{1}{n} \sum_{i} \log \frac{e^{S(f_t(q_i), f_t(d_i)) / \tau}}{\sum_{j=1}^{n} e^{S(f_t(q_i), f_t(d_j)) / \tau} + \sum_{h = 1}^{\mathcal{H}} e^{S(f_t(q_i), f_t(d_h)) / \tau}}$
    
    \IF{$t > 1$}
        \STATE $\mathcal{L}_{D} = \frac{1}{n}\sum_{i=1}^n D_c(f_t(q_i), f_{t-1}(q_i)) + D_c(f_t(d_i),f_{t-1}(d_i))$
        \STATE $\mathcal{L} = \mathcal{L}_C + \mathcal{L}_D$
    \ELSE
        \STATE $\mathcal{L} = \mathcal{L}_C$
    \ENDIF
\ENDFOR
\STATE \textcolor{gray}{\# After $f_t$ is trained, we estimate the drift vectors}
\IF{$t > 1$}
\STATE $\Delta^{t-1\rightarrow t} = \frac{1}{\mathcal{N}_t} \sum_{q \in Q_t} (f_t(q) - f_{t-1}(q))$ \quad \textcolor{gray}{\# Store $\Delta^{t-1\rightarrow t}$}
\ENDIF
\STATE \textcolor{gray}{\# Indexing of corpus documents using trained model $f_t$}
\STATE $I_t^{f_t} \leftarrow f_t(d) \; \forall \; d \in C_t$ \quad \textcolor{gray}{\# Store document embeddings in $I_t^{f_t}$}
\ENDFOR
\end{algorithmic}
\end{minipage}
\hfill
\begin{minipage}[t]{0.45\textwidth}
\begin{algorithmic}
\STATE \textbf{Retrieval Evaluation after task $t$:}
\STATE \textbf{Input:} 
\STATE $f_t$: Model from task $t$
\STATE $t' \in [1, t]$: task for evaluation
\STATE $Q_{t'}$: test query data from task $t'$
\STATE $C_{t'}$: corpus
\STATE $\{I_z^{f_z}\} \; \forall \; z \in [1,t]$: Indexed corpus embeddings
\STATE $\{\Delta^{z\rightarrow z+1}\} \; \forall \; z \in [1, t-1]$: drift vectors 
\STATE \textbf{Output:} 
\STATE $\hat{D}_{t'}$: retrieved documents corresponding to $Q_{t'}$
\vspace{2mm}
\FOR{task $t' \in [1, t]$}

\IF{$t' < t$}
    \STATE $\Delta^{t'\rightarrow t} = \sum_{j=t'}^{t-1}  \Delta^{j\rightarrow j+1}$
    \STATE $\hat{Q}_{t'}^{f_{t'}} = f_t(q_{t'}) - \Delta^{t'\rightarrow t} \; \forall \; q_{t'} \in Q_{t'}$
\ELSE
    \STATE $\hat{Q}_{t'}^{f_{t'}} = f_t(q_{t'}) \; \forall \; q_{t'} \in Q_{t'}$  \quad \textcolor{gray}{\# $t'=t$}
\ENDIF
\FOR{$\hat{q}$ in $\hat{Q}_{t'}^{f_{t'}}$}
\STATE $\mathcal{I}_i = I_{t'}^{f_{t'}}[i]$
\STATE $index = \argmax_i(S(\hat{q}, \mathcal{I}_i))$  
\STATE $\hat{d} = C_{t'}[index]$
\STATE $\hat{D}_{t'} \leftarrow \hat{d}$  \quad \textcolor{gray}{\# Store retrieved documents in $\hat{D}_{t'}$}
\ENDFOR
\ENDFOR
\end{algorithmic}
\end{minipage}
\end{small}
\end{algorithm}

After learning a new task, we estimate the drift in the embedding space from $f_{t-1}$ to $f_t$ using the queries from the training data of the current task as shown in~\cref{fig:QDC} (left). 
We project the queries $q \in Q_t$ through $f_{t-1}$ and then through $f_t$. Now that we have the queries and their corresponding embeddings from $f_{t-1}$ and $f_t$, we simply estimate the drift of queries and then average those drift vectors to obtain a single drift vector $\Delta^{t-1\rightarrow t}$ for each task $t$ as follows:
\begin{equation}
    \Delta^{t-1\rightarrow t} = \frac{1}{\mathcal{N}_t} \sum_{q \in Q_t} (f_t(q) - f_{t-1}(q)) ,
\end{equation}
where $\mathcal{N}_t$ is the number of queries in training data of task $t$. Thus, after learning a new task model $f_t$, we compute the query drift for task $t$ and store the drift vector $\Delta^{t-1\rightarrow t}$.

The drift vector from the last task can be simply added to drift vectors from previous tasks to obtain $\Delta^{t'\rightarrow t}$ as follows:
\begin{equation}
    \Delta^{t'\rightarrow t} = \Delta^{t'\rightarrow t'+1} + \dots + \Delta^{t-1 \rightarrow t} = \sum_{j=t'}^{t-1}  \Delta^{j\rightarrow j+1} ,
\end{equation}
During retrieval time, for task $t'$, we pass the queries $Q_{t'}$ through the updated model $f_t$ to obtain embeddings $Q_{t'}^{f_t}$ and then 
compensate the embeddings by subtracting the drift vector of the corresponding task $\Delta^{t'\rightarrow t}$ to project them back to the embedding space of $f_{t'}$ as illustrated in~\cref{fig:QDC} (right). For query $q_{t'} \in Q_{t'}$, we perform the query drift compensation as follows:
\begin{equation}
    \hat{Q}_{t'}^{f_{t'}} = \hat{f}_{t'}(q_{t'}) = f_t(q_{t'}) - \Delta^{t'\rightarrow t} ,
\end{equation}
where $\hat{f}_{t'}(q_{t'})$ is the set of queries which are estimated using the proposed method and is expected to be similar to $f_{t'}(q_{t'})$.
Having estimated $\hat{Q}_{t'}^{f_{t'}}$ for queries from task $t'$, we can now compare the query $\hat{q} \in \hat{Q}_{t'}^{f_{t'}}$ and indexed document embeddings $I_{t'}^{f_{t'}}$ in the same embedding space of $f_{t'}$ as follows:
\begin{equation}
    \mathcal{I}_i = I_{t'}^{f_{t'}}[i]; \quad\quad index = \argmax_i (S(\hat{q}, \mathcal{I}_i)); \quad \quad \hat{d} = C_{t'}[index] ,
\end{equation}
where $\mathcal{I}_i$ refers to embeddings from the indexed corpus $I_{t'}^{f_{t'}}$ at index $i$ and $S$ refers to the cosine similarity. Thus, we resolve the non-compatibility issue by simple vector subtraction without any re-indexing. 
We summarize the proposed approach for training and retrieval evaluation in~\cref{alg:algo}.
In the experiments, we also explore estimating multiple query drift vectors per task, but do not find this to significantly improve results.

\section{Experiments}

\textbf{Datasets.} We use 5 retrieval datasets from the BEIR benchmark~\citep{thakur2021beir} and continually train and evaluate them in task-incremental learning setup. We use the datasets in the following sequence - MS MARCO~\citep{bajaj2016ms}, NQ~\citep{kwiatkowski2019natural}, HotpotQA~\citep{yang2018hotpotqa}, FEVER~\citep{thorne2018fever} and FiQA-2018~\citep{maia201818}. We select these datasets since they have sufficient training set of query-document pairs and hence we can continually train them. The other datasets from BEIR benchmark has very little to no training data and are primarily used for zero-shot retrieval evaluation. We discuss the details of the datasets in~\cref{tab:dataset_stats}.

\textbf{Implementation.} We use the nomic embedding model~\citep{nussbaum2024nomic} with 768 dimensional feature embeddings for our experiments. We use the MTEB benchmark~\citep{muennighoff2023mteb} for evaluating the retrieval models on different tasks. We use the pre-trained model\footnote{nomic pre-trained model (\url{https://huggingface.co/nomic-ai/nomic-embed-text-v1-unsupervised})} from~\cite{nussbaum2024nomic} which is pre-trained with MLM followed by unsupervised contrastive pre-training. The nomic model is a modified version of BERT base~\citep{devlin2019bert} resulting in a 137M parameter model with 8192 sequence length. Starting with the pre-trained model, we fine-tune them continually for our experiments. Following~\cite{nussbaum2024nomic}, we use $\mathcal{H}=7$ hard negatives for each query which are mined from the corresponding dataset corpus using the gte-base model\footnote{ gte-base model (\url{https://huggingface.co/thenlper/})}~\citep{li2023towards}. For each task, we train for one epoch with a batch size of 128, learning rate of $2 \times 10^{-5}$ and weight decay of $0.01$. Similar to~\cite{nussbaum2024nomic}, we find that training for more epochs does not improve performance. We use 4 NVIDIA L40S GPUs to train the models for our experiments. While we compare the nomic model performance with other dense retrievers in~\cref{app-nomic}, the main evaluation for the continual setting is based on the pre-trained nomic model which we continually finetune on new tasks.

\textbf{Metrics.} We use the Normalised Cumulative Discount Gain (nDCG@10) metric~\citep{yining2013theoretical} for top-10 retrieved documents in our evaluations following~\cite{thakur2021beir,gunther2023jina,nussbaum2024nomic}. We also report other metrics like Recall@10 and MAP@10 (Mean Average Precision) in~\cref{app-metrics}.

\begin{table*}[t!]
    \small
    \vspace{-2mm} 
    \caption{Details of datasets used in the proposed benchmark for CDR. Details excerpted from~\cite{thakur2021beir}.}
    \centering
    \label{tab:dataset_stats}
    \resizebox{0.95\textwidth}{!}{\begin{tabular}{ l | l | l | c | c c c | c c }
        \toprule
         \multicolumn{1}{l}{\textbf{Split} ($\rightarrow$)} &
         \multicolumn{2}{c}{} &
         \multicolumn{1}{c}{\textbf{Train}}  &
         \multicolumn{3}{c}{\textbf{Test}}   &
         \multicolumn{2}{c}{\textbf{Avg.~Word Lengths}} \\
         \cmidrule(lr){4-4}
         \cmidrule(lr){5-7}
         \cmidrule(lr){8-9}
           \textbf{Task ($\downarrow$)} &\textbf{Domain ($\downarrow$)} & \textbf{Dataset ($\downarrow$)}  & \textbf{\#Pairs} & \textbf{\#Query} &  \textbf{\#Corpus} & \textbf{Avg. D~/~Q } & \textbf{Query} & \textbf{Document} \\
         \midrule
    Passage-Retrieval    & Misc. & MS MARCO & 532,761  &   6,980   &   8,841,823      & 1.1 & 5.96  & 55.98  \\ \midrule[0.05pt]
    Question             & Wikipedia  & NQ  & 132,803 & 3,452 & 2,681,468 & 1.2  & 9.16  & 78.88  \\
    Answering       & Wikipedia  & HotpotQA    & 170,000 & 7,405  & 5,233,329 & 2.0  & 17.61 & 46.30  \\
     (QA)           &Finance& FiQA-2018   & 14,166  & 648    & 57,638    & 2.6  & 10.77 & 132.32 \\ 
     \midrule
    Fact Checking         & Wikipedia  &  FEVER   & 140,085 & 6,666  & 5,416,568 & 1.2  & 8.13  & 84.76  \\                     
    \bottomrule
    \end{tabular}}
    \vspace{-1mm} 
\end{table*}

\begin{table}
\begin{center}
\caption{Performance comparison of different approaches for Continual Document Retrieval tasks. Here, we report the nDCG@10 scores for retrieval of each task. For continually trained methods, we use the latest model after training on T5 for retrieval of all tasks. We highlight the proposed QDC-based approaches in \colorbox{LightPurple}{purple}. $\dagger$: results excerpted from~\cite{thakur2021beir}.}
\resizebox{\textwidth}{!}{
\begin{tabular}{c|c|c|c|c|c|c|c}
Method & Continual & T1 - MS MARCO & T2 - NQ & T3 - Hotpot QA & T4 - FEVER & T5 - FiQA2018 & Avg. Score  \\
\hline
\rule{0pt}{2ex}
BM25$^\dagger$ & \XSolidBrush & 22.8 & 32.9 & 60.3 & 75.3 & 23.6 & 43.0 \\
Joint Training & \XSolidBrush & 39.2 & 50.7 & 71.7 & 75.1 & 33.8 & 54.1 \\
\hline
\rule{0pt}{2ex} 
FT & $\checkmark$ & 34.8 & 50.4 & 59.1 & 68.2 & 34.4 & 49.4 \\
\rowcolor{LightPurple} FT + QDC & $\checkmark$ & 38.4 & 52.5 & 67.5 & 75.0 & 34.4 & 53.5 \\
\hline
\rule{0pt}{2ex} 
FT + KD & $\checkmark$ & 37.6 & 52.6 & 65.8 & 63.7 & 34.3 & 50.8 \\
\rowcolor{LightPurple} FT + KD + QDC & $\checkmark$ & 39.3 & 54.1 & 69.3 & 73.6 & 34.3 & \textbf{54.1} \\
\hline
\rule{0pt}{2ex} 
FT (with re-indexing) & $\checkmark$ & 33.8 & 45.7 & 62.9 & 73.5 & 34.4 & 50.1 \\
FT + KD (with re-indexing) & $\checkmark$ & 34.1 & 46.7 & 64.1 & 72.6 & 34.3 & 50.4 \\

\hline
\end{tabular} 
}
\label{table_1}
\end{center}
\vspace{-3mm} 
\end{table}

\textbf{Comparison to BM25.} We compare with commonly used lexical retriever BM25~\citep{robertson2009probabilistic} in~\cref{table_1}. We observe that BM25 performs very poorly on most datasets compared to FT or joint training except on FEVER where it outperforms all other approaches.

\textbf{Impact of KD.} We show in~\cref{table_1} that embedding knowledge distillation (KD) improves the stability of the model which is evident from better performance of old tasks where FT+KD improves MS MARCO by 2.8\%, NQ by 2.2\% and HotpotQA by 6.7\% over FT. While KD improves over FT by 1.4\% on an average, we also observe that KD can affect the plasticity of the model since the performance on newer tasks are affected (FEVER drops by 4.5\%).

\textbf{Impact of QDC.} We show in~\cref{table_1} that QDC outperforms FT and FT+KD and improves the performance of all tasks after continually training on all tasks. Using QDC improves over FT across all tasks by 4.1\% on average. When used with FT+KD models, QDC improves by 3.3\% on average and outperforms all other approaches. We also evaluate the impact of QDC on the performance by evaluating on all tasks after training on each task in~\cref{table_2}. We observe poor performance of old tasks (in yellow) with naive fine-tuning. When using QDC for retrieval of old tasks, the performance improves for all tasks, thereby reducing the forgetting significantly (denoted by PD). Finally, using QDC with FT+KD models achieves the best average accuracy and least PD after each task. We present examples showing improved retrieval of documents using QDC in~\cref{app-examples}. We also discuss how to use QDC in class-incremental setting in~\cref{app-cil}.

\textbf{Comparison to joint training.} We compare the performance of continually trained model with the jointly trained model, which is trained on all five datasets at the same time in a static setting. We observe that the proposed method (FT+KD+QDC) performs similarly to the jointly trained model on average. 
Note that the joint training here considers only those hard-negatives which are mined from each dataset and are not jointly mined. In other words, we use the same hard-negatives for each dataset in both continual finetuning and joint training for fair comparison.
While the joint training performance could improve with jointly mined hard-negatives, we 
follow the standard practice of joint training~\citep{nussbaum2024nomic}. 

\begin{table}
\begin{center}
\caption{Performance of the proposed method for Continual Document Retrieval tasks after training on each task. Here, we show the retrieval performance (nDCG@10 scores) for all datasets as the model is continually trained on a new task. We denote the performance of old tasks in \colorbox{blanchedalmond}{yellow}
and the zero-shot retrieval performance of future tasks in \colorbox{celadon}{green}. The performance drop (PD) from the task in which the model is learned on a given dataset (denoted by TX) to the last task suggests the forgetting of that dataset.}
\resizebox{0.9\textwidth}{!}{
\begin{tabular}{cc|c|c|c|c|c|c}
\hline
\rule{0pt}{3ex}

& & \multicolumn{5}{c}{Training Sequence $\rightarrow$} & \\
 \cmidrule{3-8}
 \rule{0pt}{2ex} 
\multirow{7}{*}{\rotatebox[origin=c]{90}{\textbf{FT}}}
& Eval on &  T1 - MS MARCO & T2 - NQ & T3 - HotpotQA & T4 - FEVER & T5 - FiQA2018 & PD (TX - T5) \\
\cmidrule{2-8}
\rule{0pt}{2ex} 
& MS MARCO &  40.2 & \cellcolor{blanchedalmond} 38.7 & \cellcolor{blanchedalmond} 38.2 & \cellcolor{blanchedalmond} 36.3 & \cellcolor{blanchedalmond} 34.8 & 5.4 \\
& NQ & \cellcolor{celadon} 50.4 & 54.7 & \cellcolor{blanchedalmond} 51.6 & \cellcolor{blanchedalmond} 50.7 & \cellcolor{blanchedalmond} 50.4 & 4.3 \\
& HotpotQA & \cellcolor{celadon} 61.3 & \cellcolor{celadon} 56.4 & 71.5 & \cellcolor{blanchedalmond} 68.8 & \cellcolor{blanchedalmond} 59.1 & 12.4 \\
& FEVER & \cellcolor{celadon} 67.8 & \cellcolor{celadon} 58.0 & \cellcolor{celadon} 66.1 & 72.8 & \cellcolor{blanchedalmond} 68.2 & 4.6 \\
& FiQA2018 & \cellcolor{celadon} 31.4 & \cellcolor{celadon} 32.4 & \cellcolor{celadon} 29.4 & \cellcolor{celadon} 29.3 & 34.4 & - \\
\cmidrule{2-8}
\rule{0pt}{2ex} 
& Avg Acc. & 50.2 & 48.1 & 51.4 & 51.6 & 49.4 \\

\hline
\rule{0pt}{3ex} 

\multirow{7}{*}{\rotatebox[origin=c]{90}{\textbf{FT+QDC}}}

 & Eval on &  T1 - MS MARCO & T2 - NQ & T3 - HotpotQA & T4 - FEVER & T5 - FiQA2018 & PD (TX - T5) \\
\cmidrule{2-8}
\rule{0pt}{2ex} 
& MS MARCO &  40.2 & \cellcolor{blanchedalmond} 39.9 & \cellcolor{blanchedalmond} 38.5 & \cellcolor{blanchedalmond} 37.8 & \cellcolor{blanchedalmond} 38.4 & 1.8 \\
& NQ & \cellcolor{celadon} 50.4 & 54.7 & \cellcolor{blanchedalmond} 53.0 & \cellcolor{blanchedalmond} 50.8 & \cellcolor{blanchedalmond} 52.5 & 2.2 \\
& HotpotQA & \cellcolor{celadon} 61.3 & \cellcolor{celadon} 56.4 & 71.5 & \cellcolor{blanchedalmond} 70.5 & \cellcolor{blanchedalmond} 67.5 & 4.0\\
& FEVER & \cellcolor{celadon} 67.8 & \cellcolor{celadon} 58.0 & \cellcolor{celadon} 66.1 & 72.8 & \cellcolor{blanchedalmond} 75.0 & -2.2 \\
& FiQA2018 & \cellcolor{celadon} 31.4 & \cellcolor{celadon} 32.4 & \cellcolor{celadon} 29.4 & \cellcolor{celadon} 29.3 & 34.4 & - \\
\cmidrule{2-8}
\rule{0pt}{2ex} 
& Avg Acc. & 50.2 & 48.3 & 51.7  & 52.2 & 53.5 \\

\hline
\rule{0pt}{3ex} 

\multirow{7}{*}{\rotatebox[origin=c]{90}{\textbf{FT+KD+QDC}}}

 & Eval on &  T1 - MS MARCO & T2 - NQ & T3 - HotpotQA & T4 - FEVER & T5 - FiQA2018 & PD (TX - T5) \\
\cmidrule{2-8}
\rule{0pt}{2ex} 
& MS MARCO &  40.2 & \cellcolor{blanchedalmond} 40.2 & \cellcolor{blanchedalmond} 39.4 & \cellcolor{blanchedalmond} 39.0 & \cellcolor{blanchedalmond} 39.3 & 0.9 \\
& NQ & \cellcolor{celadon} 50.4 & 54.7 & \cellcolor{blanchedalmond} 54.1 & \cellcolor{blanchedalmond} 52.8 & \cellcolor{blanchedalmond} 54.1 & 0.6 \\
& HotpotQA & \cellcolor{celadon} 61.3 & \cellcolor{celadon} 58.7 & 72.7 & \cellcolor{blanchedalmond} 72.5 & \cellcolor{blanchedalmond} 69.3 & 3.4\\
& FEVER & \cellcolor{celadon} 67.8 & \cellcolor{celadon} 61.1 & \cellcolor{celadon} 67.4 & 70.5 & \cellcolor{blanchedalmond} 73.6 & -3.1 \\
& FiQA2018 & \cellcolor{celadon} 31.4 & \cellcolor{celadon} 32.7 & \cellcolor{celadon} 30.3 & \cellcolor{celadon} 29.3 & 34.3 & - \\
\cmidrule{2-8}
\rule{0pt}{2ex} 
& Avg Acc. & 50.2 & 49.5 & 52.8 & 52.8 & 54.1 \\
\hline

\end{tabular} 
}
\vspace{-15pt}
\label{table_2}
\end{center}
\end{table}

\noindent\textbf{Generalizability.} We also evaluate the zero-shot performance in~\cref{table_2} (in green) for unseen tasks and demonstrate how training on each new task affects the generalizability of the model. The zero-shot performance depends on the latest task the model is trained on. For instance, we see a drop in performance of FEVER after training on NQ while it improves after training on Hotpot QA. This could be due to high domain overlap between Hotpot QA and FEVER as shown in~\citep{thakur2021beir}. We also observe an improvement in zero-shot performance when using KD.

\begin{figure}
    \centering
    \resizebox{0.9\textwidth}{!}{
    \begin{subfigure}[b]{0.48\columnwidth}
         \centering
         \begin{subfigure}[b]{\columnwidth}
             \centering
             \includegraphics[width=\columnwidth]{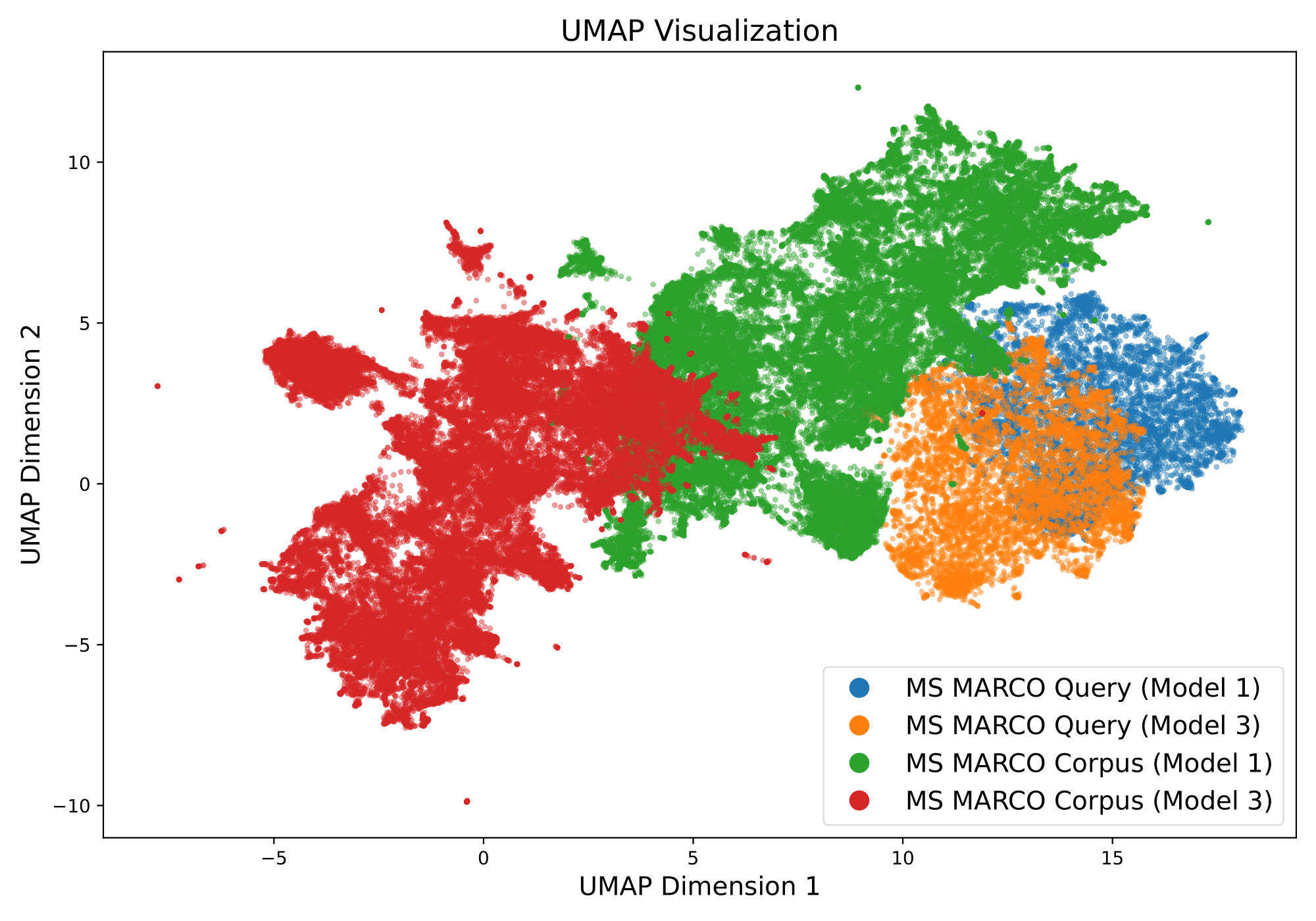}
         \end{subfigure}
    \end{subfigure}
    \hfill
    \begin{subfigure}[b]{0.41\columnwidth}
         \centering
         \begin{subfigure}[b]{\columnwidth}
             \centering
             \includegraphics[width=\columnwidth]{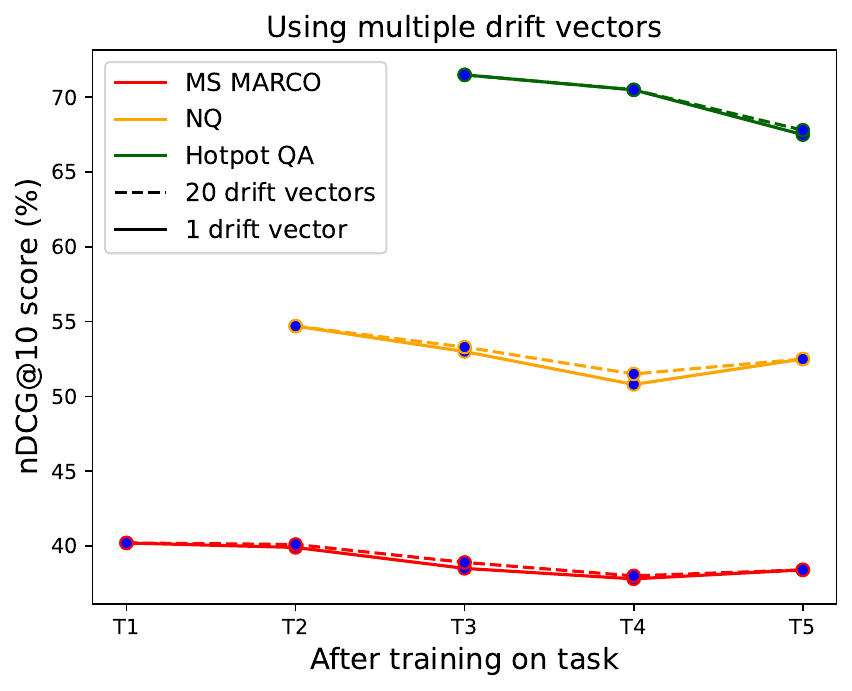}
         \end{subfigure}
    \end{subfigure}
    }
    \caption{(left) UMAP visualization of the drift in query and corpus embeddings of MS MARCO after fine-tuning on NQ and Hotpot QA (using Model after task 3). (right) Analysis of the performance when using multiple drift vectors to represent drift between embedding spaces of old and new task.}
    \label{fig:ablation}
    \vspace{-15pt}
\end{figure}

\textbf{Comparison to re-indexing.} While re-indexing was found to be effective in image retrieval~\citep{shen2020towards}, we show that in CDR, re-indexing of old task documents does not improve performance significantly and performs worse compared to FT on initial tasks like MS MARCO and NQ. The proposed method QDC significantly outperforms re-indexing approach, even with re-indexing on the model trained with embedding distillation as shown in~\cref{table_1}.
Previously, \citeauthor{wang2021continual} observed unequal drifts in image and text modalities for continual multimodal retrieval. We observe a similar phenomenon here and show in~\cref{fig:ablation} (left) that the drift in corpus document embeddings is higher compared to the drift in query embeddings for MS MARCO after training on NQ followed by Hotpot QA. Thus, the corpus document embeddings (in red) are poorly aligned with the query embeddings (in orange) in the updated embedding space after task 3. 

We attribute the poor performance of re-indexing in CDR to the unequal drift in query and document embeddings. We hypothesize that the unequal drift could be due to the difference in length of queries and documents. Queries are much shorter in length while documents could be a paragraph. From~\cref{tab:dataset_stats}, we observe that documents have a much higher average word lengths than queries (MS MARCO documents are 10 times longer than the queries on an average). This suggests that the documents could face higher forgetting or higher embedding drift compared to queries.

In the case of multi-modal retrieval~\citep{wang2021continual}, the unequal drift arises from the modality gap at initialization which is preserved during the contrastive training approach (as discussed in \cite{liang2022mind} for vision-language models). A recent work~\citep{schrodi2025two} extensively analyzed and presented that the gap between modalities actually arises from the information imbalance in the two modalities where one modality (visual) has more information than the other one (text). We think that this explanation could also be applicable to our setting despite that we only have a single modality, since there is a significant information imbalance between queries and corpus. 
The imbalance in information is caused by the difference in the lengths of queries and corpus where queries are usually much shorter in length and the corpus are much longer and more detailed. 

\begin{wraptable}{r}{0.35\textwidth}
\centering
\vspace{-10pt}
\caption{Average drift values for Query and Corpus across text lengths.}
\begin{tabular}{@{}lccc@{}}
\toprule
\textbf{} & \textbf{Short} & \textbf{Medium} & \textbf{Long} \\
\midrule
\textbf{Query} & 0.0340 & 0.0375 & 0.0393 \\
\textbf{Corpus} & 0.0559 & 0.0573 & 0.0635 \\
\bottomrule
\end{tabular}
\vspace{-10pt}
\label{avg_drift}
\end{wraptable}

We analyze in~\cref{avg_drift} how the queries and corpus of different lengths of the first task MS Marco drift after training on NQ. We divide the queries and corpus into three groups (short, medium and long) based on lengths and report the average of the cosine drift in each group. We observe that corpus documents having more information drift more than the queries as also seen in~\cref{fig:ablation} (left). We observe that within the queries, the medium-sized and longer queries drift more on average than the shorter ones. Similarly, among the corpus documents, we see that longer documents drift more than the short ones. This confirms the correlation between the length of the query or document and the drift. 

\textbf{Enabling compatibility with QDC.} There are two ways of maintaining query and corpus compatibility by keeping them in the same embedding space, one by re-indexing (bringing documents forward to the new space) and the other by projecting queries back to the old embedding space. While re-indexing solves the embedding space alignment problem and brings query and corpus in the same space, it does not maintain good discriminative power for the old task corpus documents since it uses the updated model which is not the best model for the old task. On the other hand, we preserve the discriminative capabilities of the model to encode the corpus in QDC since we use the documents indexed with the old model as shown in~\cref{properties}.

So, despite adding more computation costs, FT with re-indexing uses the updated model with reduced discriminative capabilities for both queries and the corpus of the old task. While QDC still uses the best model for the old task for indexing the documents and the drift compensated queries solves the alignment problem. Note that the corpus embeddings play a more important role here since a single query embedding is searched across millions of corpus documents. So, it is more important to preserve the corpus embedding space for old tasks and the best way is to use the old model indexed embeddings.

\textbf{QDC with multiple drift vectors.} In the proposed method, we consider a single drift vector for each task transition. One could also estimate multiple drift vectors in the embedding space. This could be done by dividing the embedding space into $k$ clusters with corresponding centroids $P_k$ and estimating a drift vector for each of these centroids $\Delta^{t-1\rightarrow t}_k$ by averaging the drift of queries belonging to each cluster. During retrieval, the queries $q_{t'}$ could be assigned to the closest centroid $k'$ and then compensated with the drift vector of the centroid $\Delta^{t'\rightarrow t}_{k'}$ as follows:
\begin{equation}
    k' = \argmax_k S(P_k, f_t(q_{t'})) ; \quad \quad
    \hat{f}_{t'}(q_{t'}) = f_t(q_{t'}) - \Delta^{t'\rightarrow t}_{k'} \; 
\end{equation}
where $k'$ is the closest cluster to the query embedding. Using multiple drift vectors involves storing the cluster centroids and the drift vectors for each centroid. 

In our experiments in~\cref{fig:ablation} (right), we empirically demonstrate that using a single drift vector is effective to estimate the query drift and estimating multiple drift vectors for each task transition does not improve the performance significantly. Using 20 drift vectors improves the performance of MS MARCO by 0.4\% after T3, NQ by 0.7\% after T4, Hotpot QA by 0.3\% after T5, while achieving the same performance for FEVER. Based on these observations, we advocate using a single drift vector for each task which achieves similar performance with simpler and faster retrieval since it does not need to find the closest cluster centroid for drift compensation. 

\begin{table}[t]
\centering
\caption{Comparison of methods and their properties across tasks.}
\resizebox{0.8\textwidth}{!}{
\begin{tabular}{@{}cccccc@{}}
\toprule
Task & Method & Query Embedding & Corpus Embedding & Compatibility & Discriminative Power for T1 \\
\midrule
T1 & FT                    & T1 & T1 & \Checkmark & \Checkmark \\
\midrule
T2 & FT                    & T2 & T1 & \XSolidBrush      & \Checkmark \\
T2 & FT with re-indexing   & T2 & T2 & \Checkmark & \XSolidBrush      \\
\rowcolor{LightPurple}
T2 & FT+QDC                & T1 & T1 & \Checkmark & \Checkmark \\
\bottomrule
\end{tabular}
}
\label{properties}
\end{table}

\section{Conclusion}
In this work, we study how continually training embedding models on query-document pairs from new datasets over time could affect the retrieval performance across all seen tasks. We observe forgetting on old tasks in CDR and show that using knowledge distillation on the query and document embeddings can reduce the forgetting. We discuss the issue of non-compatibility between query and indexed corpus embeddings due to \textit{embedding drift} after the embedding model is updated on a new task. We propose a novel method to avoid this issue by estimating the drift of queries from old to new embedding space and then compensating the estimated drift to project the queries to old embedding space at test time. This enables compatibility since the indexed embeddings were extracted from the old model and thus we compute similarities for retrieval with queries and corpus in the same embedding space. We establish a continual training benchmark with five large-scale datasets and demonstrate that the proposed QDC approach outperforms other approaches. We also show that re-indexing based approach does not perform well despite being very expensive. 

We believe that enabling compatibility in continually trained retrieval embedding models will benefit several practical document retrieval applications like RAG systems where the retriever embedding model could be continually updated to add new knowledge over time. We hope that our approach and findings will encourage further research and more extensive benchmarks on continual document retrieval.

\paragraph{Acknowledgements.} We acknowledge projects PID2022-143257NB-I00, financed by MCIN/AEI/10.13039/501100011033 and FSE+, funding by the European Union ELLIOT project, and the Generalitat de Catalunya CERCA Program. Dipam Goswami acknowledges travel support from ELISE (GA no 951847). Bartłomiej Twardowski acknowledges the grant RYC2021-032765-I and National Centre of Science (NCN, Poland) Grant No. 2023/51/D/ST6/02846. Liying Wang acknowledges financial support from China Scholarship Council.

\bibliography{collas2025_conference}

\begin{thebibliography}{68}
\providecommand{\natexlab}[1]{#1}
\providecommand{\url}[1]{\texttt{#1}}
\expandafter\ifx\csname urlstyle\endcsname\relax
  \providecommand{\doi}[1]{doi: #1}\else
  \providecommand{\doi}{doi: \begingroup \urlstyle{rm}\Url}\fi

\bibitem[Bajaj et~al.(2016)Bajaj, Campos, Craswell, Deng, Gao, Liu, Majumder, McNamara, Mitra, Nguyen, et~al.]{bajaj2016ms}
Payal Bajaj, Daniel Campos, Nick Craswell, Li~Deng, Jianfeng Gao, Xiaodong Liu, Rangan Majumder, Andrew McNamara, Bhaskar Mitra, Tri Nguyen, et~al.
\newblock Ms marco: A human generated machine reading comprehension dataset.
\newblock \emph{arXiv preprint arXiv:1611.09268}, 2016.

\bibitem[Belouadah \& Popescu(2019)Belouadah and Popescu]{belouadah2019il2m}
Eden Belouadah and Adrian Popescu.
\newblock Il2m: Class incremental learning with dual memory.
\newblock In \emph{International Conference on Computer Vision (ICCV)}, 2019.

\bibitem[Berger et~al.(2000)Berger, Caruana, Cohn, Freitag, and Mittal]{berger2000bridging}
Adam Berger, Rich Caruana, David Cohn, Dayne Freitag, and Vibhu Mittal.
\newblock Bridging the lexical chasm: statistical approaches to answer-finding.
\newblock In \emph{Proceedings of the 23rd annual international ACM SIGIR conference on Research and development in information retrieval}, pp.\  192--199, 2000.

\bibitem[Biondi et~al.(2023)Biondi, Pernici, Bruni, Mugnai, and Bimbo]{biondi2023cl2r}
Niccol{\'o} Biondi, Federico Pernici, Matteo Bruni, Daniele Mugnai, and Alberto~Del Bimbo.
\newblock Cl2r: Compatible lifelong learning representations.
\newblock \emph{ACM Transactions on Multimedia Computing, Communications and Applications}, 18\penalty0 (2s):\penalty0 1--22, 2023.

\bibitem[Biondi et~al.(2024)Biondi, Pernici, Ricci, and Del~Bimbo]{biondi2024stationary}
Niccol{\`o} Biondi, Federico Pernici, Simone Ricci, and Alberto Del~Bimbo.
\newblock Stationary representations: Optimally approximating compatibility and implications for improved model replacements.
\newblock In \emph{Proceedings of the IEEE/CVF Conference on Computer Vision and Pattern Recognition}, pp.\  28793--28804, 2024.

\bibitem[Bornschein et~al.(2023)Bornschein, Galashov, Hemsley, Rannen-Triki, Chen, Chaudhry, He, Douillard, Caccia, Feng, et~al.]{bornschein2023nevis}
Jorg Bornschein, Alexandre Galashov, Ross Hemsley, Amal Rannen-Triki, Yutian Chen, Arslan Chaudhry, Xu~Owen He, Arthur Douillard, Massimo Caccia, Qixuan Feng, et~al.
\newblock Nevis' 22: A stream of 100 tasks sampled from 30 years of computer vision research.
\newblock \emph{Journal of Machine Learning Research}, 24\penalty0 (308):\penalty0 1--77, 2023.

\bibitem[Cai et~al.(2023)Cai, Bi, Fan, Guo, Chen, and Cheng]{cai_L2R_2023_ACMIKM}
Yinqiong Cai, Keping Bi, Yixing Fan, Jiafeng Guo, Wei Chen, and Xueqi Cheng.
\newblock L2r: Lifelong learning for first-stage retrieval with backward-compatible representations.
\newblock In \emph{Proceedings of the 32nd ACM International Conference on Information and Knowledge Management}, pp.\  183–192, 2023.

\bibitem[Cer et~al.(2018)Cer, Yang, Kong, Hua, Limtiaco, John, Constant, Guajardo-Cespedes, Yuan, Tar, et~al.]{cer2018universal}
Daniel Cer, Yinfei Yang, Sheng-yi Kong, Nan Hua, Nicole Limtiaco, Rhomni~St John, Noah Constant, Mario Guajardo-Cespedes, Steve Yuan, Chris Tar, et~al.
\newblock Universal sentence encoder for english.
\newblock In \emph{Proceedings of the 2018 conference on empirical methods in natural language processing: system demonstrations}, pp.\  169--174, 2018.

\bibitem[De~Lange et~al.(2021)De~Lange, Aljundi, Masana, Parisot, Jia, Leonardis, Slabaugh, and Tuytelaars]{de2021continual}
Matthias De~Lange, Rahaf Aljundi, Marc Masana, Sarah Parisot, Xu~Jia, Ale{\v{s}} Leonardis, Gregory Slabaugh, and Tinne Tuytelaars.
\newblock A continual learning survey: Defying forgetting in classification tasks.
\newblock \emph{Transactions on Pattern Analysis and Machine Intelligence (T-PAMI)}, 2021.

\bibitem[Devlin et~al.(2019)Devlin, Chang, Lee, and Toutanova]{devlin2019bert}
Jacob Devlin, Ming-Wei Chang, Kenton Lee, and Kristina Toutanova.
\newblock Bert: Pre-training of deep bidirectional transformers for language understanding.
\newblock In \emph{Proceedings of the 2019 conference of the North American chapter of the association for computational linguistics: human language technologies, volume 1 (long and short papers)}, pp.\  4171--4186, 2019.

\bibitem[Douillard et~al.(2020)Douillard, Cord, Ollion, Robert, and Valle]{douillard2020podnet}
Arthur Douillard, Matthieu Cord, Charles Ollion, Thomas Robert, and Eduardo Valle.
\newblock Podnet: Pooled outputs distillation for small-tasks incremental learning.
\newblock In \emph{European Conference on Computer Vision (ECCV)}, 2020.

\bibitem[Gerald \& Soulier(2022)Gerald and Soulier]{GeraldSoulier_CLforLTSinNIR_2022_AIR}
Thomas Gerald and Laure Soulier.
\newblock Continual learning of long topic sequences in neural information retrieval.
\newblock In \emph{Advances in Information Retrieval}, pp.\  244--259, 2022.

\bibitem[Gillick et~al.(2018)Gillick, Presta, and Tomar]{gillick2018end}
Daniel Gillick, Alessandro Presta, and Gaurav~Singh Tomar.
\newblock End-to-end retrieval in continuous space.
\newblock \emph{arXiv preprint arXiv:1811.08008}, 2018.

\bibitem[Gomez-Villa et~al.(2025)Gomez-Villa, Goswami, Wang, Bagdanov, Twardowski, and van~de Weijer]{Alex_LDC_2024_ECCV}
Alex Gomez-Villa, Dipam Goswami, Kai Wang, Andrew~D. Bagdanov, Bartlomiej Twardowski, and Joost van~de Weijer.
\newblock Exemplar-free continual representation learning via learnable drift compensation.
\newblock In \emph{Computer Vision -- ECCV 2024}, pp.\  473--490, 2025.

\bibitem[Goswami et~al.(2023)Goswami, Liu, Twardowski, and van~de Weijer]{goswami2023fecam}
Dipam Goswami, Yuyang Liu, Bart{\l}omiej Twardowski, and Joost van~de Weijer.
\newblock Fecam: Exploiting the heterogeneity of class distributions in exemplar-free continual learning.
\newblock In \emph{Thirty-seventh Conference on Neural Information Processing Systems}, 2023.

\bibitem[Goswami et~al.(2024)Goswami, Soutif-Cormerais, Liu, Kamath, Twardowski, and van~de Weijer]{Goswami_ADC_2024_CVPR}
Dipam Goswami, Albin Soutif-Cormerais, Yuyang Liu, Sandesh Kamath, Bartlomiej Twardowski, and Joost van~de Weijer.
\newblock Resurrecting old classes with new data for exemplar-free continual learning.
\newblock In \emph{Proceedings of the IEEE/CVF Conference on Computer Vision and Pattern Recognition (CVPR)}, pp.\  28525--28534, 2024.

\bibitem[G{\"u}nther et~al.(2023)G{\"u}nther, Ong, Mohr, Abdessalem, Abel, Akram, Guzman, Mastrapas, Sturua, Wang, et~al.]{gunther2023jina}
Michael G{\"u}nther, Jackmin Ong, Isabelle Mohr, Alaeddine Abdessalem, Tanguy Abel, Mohammad~Kalim Akram, Susana Guzman, Georgios Mastrapas, Saba Sturua, Bo~Wang, et~al.
\newblock Jina embeddings 2: 8192-token general-purpose text embeddings for long documents.
\newblock \emph{arXiv preprint arXiv:2310.19923}, 2023.

\bibitem[Hou et~al.(2025)Hou, Cosma, and Finke]{hou2025advancing}
Jingrui Hou, Georgina Cosma, and Axel Finke.
\newblock Advancing continual lifelong learning in neural information retrieval: definition, dataset, framework, and empirical evaluation.
\newblock \emph{Information Sciences}, 687:\penalty0 121368, 2025.

\bibitem[Izacard et~al.(2021)Izacard, Caron, Hosseini, Riedel, Bojanowski, Joulin, and Grave]{izacard2021unsupervised}
Gautier Izacard, Mathilde Caron, Lucas Hosseini, Sebastian Riedel, Piotr Bojanowski, Armand Joulin, and Edouard Grave.
\newblock Unsupervised dense information retrieval with contrastive learning.
\newblock \emph{arXiv preprint arXiv:2112.09118}, 2021.

\bibitem[Izacard et~al.(2023)Izacard, Lewis, Lomeli, Hosseini, Petroni, Schick, Dwivedi-Yu, Joulin, Riedel, and Grave]{izacard2023atlas}
Gautier Izacard, Patrick Lewis, Maria Lomeli, Lucas Hosseini, Fabio Petroni, Timo Schick, Jane Dwivedi-Yu, Armand Joulin, Sebastian Riedel, and Edouard Grave.
\newblock Atlas: Few-shot learning with retrieval augmented language models.
\newblock \emph{Journal of Machine Learning Research}, 24\penalty0 (251):\penalty0 1--43, 2023.

\bibitem[Karpukhin et~al.(2020)Karpukhin, O{\u{g}}uz, Min, Lewis, Wu, Edunov, Chen, and Yih]{karpukhin2020dense}
Vladimir Karpukhin, Barlas O{\u{g}}uz, Sewon Min, Patrick Lewis, Ledell Wu, Sergey Edunov, Danqi Chen, and Wen~Tau Yih.
\newblock Dense passage retrieval for open-domain question answering.
\newblock In \emph{2020 Conference on Empirical Methods in Natural Language Processing, EMNLP 2020}, pp.\  6769--6781. Association for Computational Linguistics (ACL), 2020.

\bibitem[Kemker et~al.(2018)Kemker, McClure, Abitino, Hayes, and Kanan]{kemker2018measuring}
Ronald Kemker, Marc McClure, Angelina Abitino, Tyler Hayes, and Christopher Kanan.
\newblock Measuring catastrophic forgetting in neural networks.
\newblock In \emph{Proceedings of the AAAI conference on artificial intelligence}, 2018.

\bibitem[Khattab \& Zaharia(2020)Khattab and Zaharia]{khattab2020colbert}
Omar Khattab and Matei Zaharia.
\newblock Colbert: Efficient and effective passage search via contextualized late interaction over bert.
\newblock In \emph{Proceedings of the 43rd International ACM SIGIR conference on research and development in Information Retrieval}, pp.\  39--48, 2020.

\bibitem[Kim et~al.(2025)Kim, Xiao, Konishi, Ke, and Liu]{kim2025open}
Gyuhak Kim, Changnan Xiao, Tatsuya Konishi, Zixuan Ke, and Bing Liu.
\newblock Open-world continual learning: Unifying novelty detection and continual learning.
\newblock \emph{Artificial Intelligence}, 338:\penalty0 104237, 2025.

\bibitem[Kirkpatrick et~al.(2017)Kirkpatrick, Pascanu, Rabinowitz, Veness, Desjardins, Rusu, Milan, Quan, Ramalho, Grabska-Barwinska, et~al.]{kirkpatrick2017overcoming}
James Kirkpatrick, Razvan Pascanu, Neil Rabinowitz, Joel Veness, Guillaume Desjardins, Andrei~A Rusu, Kieran Milan, John Quan, Tiago Ramalho, Agnieszka Grabska-Barwinska, et~al.
\newblock Overcoming catastrophic forgetting in neural networks.
\newblock \emph{Proceedings of the National Academy of Sciences (PNAS)}, 2017.

\bibitem[Kishore et~al.(2023)Kishore, Wan, Lovelace, Artzi, and Weinberger]{kishore2023incdsi}
Varsha Kishore, Chao Wan, Justin Lovelace, Yoav Artzi, and Kilian~Q Weinberger.
\newblock Incdsi: Incrementally updatable document retrieval.
\newblock In \emph{International conference on machine learning}, pp.\  17122--17134. PMLR, 2023.

\bibitem[Kwiatkowski et~al.(2019)Kwiatkowski, Palomaki, Redfield, Collins, Parikh, Alberti, Epstein, Polosukhin, Devlin, Lee, et~al.]{kwiatkowski2019natural}
Tom Kwiatkowski, Jennimaria Palomaki, Olivia Redfield, Michael Collins, Ankur Parikh, Chris Alberti, Danielle Epstein, Illia Polosukhin, Jacob Devlin, Kenton Lee, et~al.
\newblock Natural questions: a benchmark for question answering research.
\newblock \emph{Transactions of the Association for Computational Linguistics}, 2019.

\bibitem[Lewis et~al.(2020)Lewis, Perez, Piktus, Petroni, Karpukhin, Goyal, K{\"u}ttler, Lewis, Yih, Rockt{\"a}schel, et~al.]{lewis2020retrieval}
Patrick Lewis, Ethan Perez, Aleksandra Piktus, Fabio Petroni, Vladimir Karpukhin, Naman Goyal, Heinrich K{\"u}ttler, Mike Lewis, Wen-tau Yih, Tim Rockt{\"a}schel, et~al.
\newblock Retrieval-augmented generation for knowledge-intensive nlp tasks.
\newblock \emph{Advances in Neural Information Processing Systems}, 33:\penalty0 9459--9474, 2020.

\bibitem[Li et~al.(2023)Li, Zhang, Zhang, Long, Xie, and Zhang]{li2023towards}
Zehan Li, Xin Zhang, Yanzhao Zhang, Dingkun Long, Pengjun Xie, and Meishan Zhang.
\newblock Towards general text embeddings with multi-stage contrastive learning.
\newblock \emph{arXiv preprint arXiv:2308.03281}, 2023.

\bibitem[Li \& Hoiem(2018)Li and Hoiem]{li_LWF_2018_PAMI}
Zhizhong Li and Derek Hoiem.
\newblock Learning without forgetting.
\newblock \emph{IEEE Transactions on Pattern Analysis and Machine Intelligence}, 40\penalty0 (12):\penalty0 2935--2947, 2018.

\bibitem[Liang et~al.(2020)Liang, Xu, Shakeri, Santos, Nallapati, Huang, and Xiang]{liang2020embedding}
Davis Liang, Peng Xu, Siamak Shakeri, Cicero Nogueira~dos Santos, Ramesh Nallapati, Zhiheng Huang, and Bing Xiang.
\newblock Embedding-based zero-shot retrieval through query generation.
\newblock \emph{arXiv preprint arXiv:2009.10270}, 2020.

\bibitem[Liang et~al.(2022)Liang, Zhang, Kwon, Yeung, and Zou]{liang2022mind}
Victor~Weixin Liang, Yuhui Zhang, Yongchan Kwon, Serena Yeung, and James~Y Zou.
\newblock Mind the gap: Understanding the modality gap in multi-modal contrastive representation learning.
\newblock \emph{Advances in Neural Information Processing Systems}, 35:\penalty0 17612--17625, 2022.

\bibitem[Lov{\'o}n-Melgarejo et~al.(2021)Lov{\'o}n-Melgarejo, Soulier, Pinel-Sauvagnat, and Tamine]{Lov_CSinNR_2021_AIR}
Jes{\'u}s Lov{\'o}n-Melgarejo, Laure Soulier, Karen Pinel-Sauvagnat, and Lynda Tamine.
\newblock Studying catastrophic forgetting in neural ranking models.
\newblock In \emph{Advances in Information Retrieval}, pp.\  375--390, 2021.

\bibitem[Luan et~al.(2021)Luan, Eisenstein, Toutanova, and Collins]{luan2021sparse}
Yi~Luan, Jacob Eisenstein, Kristina Toutanova, and Michael Collins.
\newblock Sparse, dense, and attentional representations for text retrieval.
\newblock \emph{Transactions of the Association for Computational Linguistics}, 9:\penalty0 329--345, 2021.

\bibitem[Magistri et~al.(2024)Magistri, Trinci, Soutif, van~de Weijer, and Bagdanov]{magistrielastic}
Simone Magistri, Tomaso Trinci, Albin Soutif, Joost van~de Weijer, and Andrew~D Bagdanov.
\newblock Elastic feature consolidation for cold start exemplar-free incremental learning.
\newblock In \emph{The Twelfth International Conference on Learning Representations}, 2024.

\bibitem[Maia et~al.(2018)Maia, Handschuh, Freitas, Davis, McDermott, Zarrouk, and Balahur]{maia201818}
Macedo Maia, Siegfried Handschuh, Andr{\'e} Freitas, Brian Davis, Ross McDermott, Manel Zarrouk, and Alexandra Balahur.
\newblock Www'18 open challenge: financial opinion mining and question answering.
\newblock In \emph{Companion proceedings of the the web conference 2018}, pp.\  1941--1942, 2018.

\bibitem[Mallya \& Lazebnik(2018)Mallya and Lazebnik]{mallya2018packnet}
Arun Mallya and Svetlana Lazebnik.
\newblock Packnet: Adding multiple tasks to a single network by iterative pruning.
\newblock In \emph{Proceedings of the IEEE conference on Computer Vision and Pattern Recognition}, pp.\  7765--7773, 2018.

\bibitem[Masana et~al.(2022)Masana, Liu, Twardowski, Menta, Bagdanov, and van~de Weijer]{masana2020class}
Marc Masana, Xialei Liu, Bartlomiej Twardowski, Mikel Menta, Andrew~D Bagdanov, and Joost van~de Weijer.
\newblock Class-incremental learning: survey and performance evaluation.
\newblock \emph{Transactions on Pattern Analysis and Machine Intelligence (T-PAMI)}, 2022.

\bibitem[McCloskey \& Cohen(1989)McCloskey and Cohen]{mccloskey1989catastrophic}
Michael McCloskey and Neal~J Cohen.
\newblock Catastrophic interference in connectionist networks: The sequential learning problem.
\newblock In \emph{Psychology of learning and motivation}. Elsevier, 1989.

\bibitem[Mehta et~al.(2023)Mehta, Gupta, Tay, Dehghani, Tran, Rao, Najork, Strubell, and Metzler]{mehta2023dsi++}
Sanket~Vaibhav Mehta, Jai Gupta, Yi~Tay, Mostafa Dehghani, Vinh~Q Tran, Jinfeng Rao, Marc Najork, Emma Strubell, and Donald Metzler.
\newblock Dsi++: Updating transformer memory with new documents.
\newblock In \emph{Proceedings of the 2023 Conference on Empirical Methods in Natural Language Processing}, pp.\  8198--8213, 2023.

\bibitem[Muennighoff(2022)]{muennighoff2022sgpt}
Niklas Muennighoff.
\newblock Sgpt: Gpt sentence embeddings for semantic search.
\newblock \emph{arXiv preprint arXiv:2202.08904}, 2022.

\bibitem[Muennighoff et~al.(2023)Muennighoff, Tazi, Magne, and Reimers]{muennighoff2023mteb}
Niklas Muennighoff, Nouamane Tazi, Loic Magne, and Nils Reimers.
\newblock Mteb: Massive text embedding benchmark.
\newblock In \emph{Proceedings of the 17th Conference of the European Chapter of the Association for Computational Linguistics}, pp.\  2014--2037, 2023.

\bibitem[Nussbaum et~al.(2024)Nussbaum, Morris, Duderstadt, and Mulyar]{nussbaum2024nomic}
Zach Nussbaum, John~X Morris, Brandon Duderstadt, and Andriy Mulyar.
\newblock Nomic embed: Training a reproducible long context text embedder.
\newblock \emph{arXiv preprint arXiv:2402.01613}, 2024.

\bibitem[Oord et~al.(2018)Oord, Li, and Vinyals]{oord2018representation}
Aaron van~den Oord, Yazhe Li, and Oriol Vinyals.
\newblock Representation learning with contrastive predictive coding.
\newblock \emph{arXiv preprint arXiv:1807.03748}, 2018.

\bibitem[Petit et~al.(2023)Petit, Popescu, Schindler, Picard, and Delezoide]{petit2023fetril}
Gr{\'e}goire Petit, Adrian Popescu, Hugo Schindler, David Picard, and Bertrand Delezoide.
\newblock Fetril: Feature translation for exemplar-free class-incremental learning.
\newblock In \emph{Winter Conference on Applications of Computer Vision (WACV)}, 2023.

\bibitem[Ram et~al.(2023)Ram, Levine, Dalmedigos, Muhlgay, Shashua, Leyton-Brown, and Shoham]{ram2023context}
Ori Ram, Yoav Levine, Itay Dalmedigos, Dor Muhlgay, Amnon Shashua, Kevin Leyton-Brown, and Yoav Shoham.
\newblock In-context retrieval-augmented language models.
\newblock \emph{Transactions of the Association for Computational Linguistics}, 11:\penalty0 1316--1331, 2023.

\bibitem[Ramanujan et~al.(2022)Ramanujan, Vasu, Farhadi, Tuzel, and Pouransari]{ramanujan2022forward}
Vivek Ramanujan, Pavan Kumar~Anasosalu Vasu, Ali Farhadi, Oncel Tuzel, and Hadi Pouransari.
\newblock Forward compatible training for large-scale embedding retrieval systems.
\newblock In \emph{Proceedings of the IEEE/CVF Conference on Computer Vision and Pattern Recognition}, pp.\  19386--19395, 2022.

\bibitem[Rebuffi et~al.(2017)Rebuffi, Kolesnikov, Sperl, and Lampert]{Rebuffi_iCARL_2017_CVPR}
Sylvestre-Alvise Rebuffi, Alexander Kolesnikov, Georg Sperl, and Christoph~H. Lampert.
\newblock icarl: Incremental classifier and representation learning.
\newblock In \emph{Proceedings of the IEEE Conference on Computer Vision and Pattern Recognition (CVPR)}, July 2017.

\bibitem[Robertson et~al.(2009)Robertson, Zaragoza, et~al.]{robertson2009probabilistic}
Stephen Robertson, Hugo Zaragoza, et~al.
\newblock The probabilistic relevance framework: Bm25 and beyond.
\newblock \emph{Foundations and Trends{\textregistered} in Information Retrieval}, 3\penalty0 (4):\penalty0 333--389, 2009.

\bibitem[Schrodi et~al.(2025)Schrodi, Hoffmann, Argus, Fischer, and Brox]{schrodi2025two}
Simon Schrodi, David~T. Hoffmann, Max Argus, Volker Fischer, and Thomas Brox.
\newblock Two effects, one trigger: On the modality gap, object bias, and information imbalance in contrastive vision-language models.
\newblock In \emph{The Thirteenth International Conference on Learning Representations}, 2025.
\newblock URL \url{https://openreview.net/forum?id=uAFHCZRmXk}.

\bibitem[Serra et~al.(2018)Serra, Suris, Miron, and Karatzoglou]{serra2018overcoming}
Joan Serra, Didac Suris, Marius Miron, and Alexandros Karatzoglou.
\newblock Overcoming catastrophic forgetting with hard attention to the task.
\newblock In \emph{International conference on machine learning}, pp.\  4548--4557. PMLR, 2018.

\bibitem[Shen et~al.(2020)Shen, Xiong, Xia, and Soatto]{shen2020towards}
Yantao Shen, Yuanjun Xiong, Wei Xia, and Stefano Soatto.
\newblock Towards backward-compatible representation learning.
\newblock In \emph{Proceedings of the IEEE/CVF Conference on Computer Vision and Pattern Recognition}, pp.\  6368--6377, 2020.

\bibitem[Smith et~al.(2023)Smith, Tian, Halbe, Hsu, and Kira]{smith2023closer}
James~Seale Smith, Junjiao Tian, Shaunak Halbe, Yen-Chang Hsu, and Zsolt Kira.
\newblock A closer look at rehearsal-free continual learning.
\newblock In \emph{Proceedings of the IEEE/CVF conference on computer vision and pattern recognition}, pp.\  2410--2420, 2023.

\bibitem[Thakur et~al.(2021)Thakur, Reimers, R{\"u}ckl{\'e}, Srivastava, and Gurevych]{thakur2021beir}
Nandan Thakur, Nils Reimers, Andreas R{\"u}ckl{\'e}, Abhishek Srivastava, and Iryna Gurevych.
\newblock {BEIR}: A heterogeneous benchmark for zero-shot evaluation of information retrieval models.
\newblock In \emph{Thirty-fifth Conference on Neural Information Processing Systems Datasets and Benchmarks Track (Round 2)}, 2021.

\bibitem[Thorne et~al.(2018)Thorne, Vlachos, Christodoulopoulos, and Mittal]{thorne2018fever}
James Thorne, Andreas Vlachos, Christos Christodoulopoulos, and Arpit Mittal.
\newblock Fever: a large-scale dataset for fact extraction and verification.
\newblock In \emph{Proceedings of the 2018 Conference of the North American Chapter of the Association for Computational Linguistics: Human Language Technologies, Volume 1 (Long Papers)}, pp.\  809--819, 2018.

\bibitem[Van~de Ven \& Tolias(2019)Van~de Ven and Tolias]{van2019three}
Gido~M Van~de Ven and Andreas~S Tolias.
\newblock Three scenarios for continual learning.
\newblock \emph{arXiv preprint arXiv:1904.07734}, 2019.

\bibitem[Vaswani et~al.(2017)Vaswani, Shazeer, Parmar, Uszkoreit, Jones, and Gomez]{vaswani2017u}
Ashish Vaswani, Noam Shazeer, Niki Parmar, Jakob Uszkoreit, Llion Jones, and Aidan~N Gomez.
\newblock Attention is all you need.
\newblock \emph{Advances in neural information processing systems}, 30:\penalty0 5998--6008, 2017.

\bibitem[Verwimp et~al.(2024)Verwimp, Aljundi, Ben-David, Bethge, Cossu, Gepperth, Hayes, H{\"u}llermeier, Kanan, Kudithipudi, et~al.]{verwimpcontinual}
Eli Verwimp, Rahaf Aljundi, Shai Ben-David, Matthias Bethge, Andrea Cossu, Alexander Gepperth, Tyler~L Hayes, Eyke H{\"u}llermeier, Christopher Kanan, Dhireesha Kudithipudi, et~al.
\newblock Continual learning: Applications and the road forward.
\newblock \emph{Transactions on Machine Learning Research}, 2024.

\bibitem[Wan et~al.(2022)Wan, Chen, Wu, and Chen]{wan2022continual}
Timmy~ST Wan, Jun-Cheng Chen, Tzer-Yi Wu, and Chu-Song Chen.
\newblock Continual learning for visual search with backward consistent feature embedding.
\newblock In \emph{Proceedings of the IEEE/CVF Conference on Computer Vision and Pattern Recognition}, pp.\  16702--16711, 2022.

\bibitem[Wang et~al.(2021)Wang, Herranz, and van~de Weijer]{wang2021continual}
Kai Wang, Luis Herranz, and Joost van~de Weijer.
\newblock Continual learning in cross-modal retrieval.
\newblock In \emph{Proceedings of the IEEE/CVF conference on computer vision and pattern recognition}, pp.\  3628--3638, 2021.

\bibitem[Wang et~al.(2022)Wang, Yang, Huang, Jiao, Yang, Jiang, Majumder, and Wei]{wang2022text}
Liang Wang, Nan Yang, Xiaolong Huang, Binxing Jiao, Linjun Yang, Daxin Jiang, Rangan Majumder, and Furu Wei.
\newblock Text embeddings by weakly-supervised contrastive pre-training.
\newblock \emph{arXiv preprint arXiv:2212.03533}, 2022.

\bibitem[Wang et~al.(2024)Wang, Zhang, Su, and Zhu]{wang2024comprehensive}
Liyuan Wang, Xingxing Zhang, Hang Su, and Jun Zhu.
\newblock A comprehensive survey of continual learning: theory, method and application.
\newblock \emph{IEEE Transactions on Pattern Analysis and Machine Intelligence}, 2024.

\bibitem[Xiong et~al.(2021)Xiong, Xiong, Li, Tang, Liu, Bennett, Ahmed, and Overwijk]{xiongapproximate}
Lee Xiong, Chenyan Xiong, Ye~Li, Kwok-Fung Tang, Jialin Liu, Paul~N Bennett, Junaid Ahmed, and Arnold Overwijk.
\newblock Approximate nearest neighbor negative contrastive learning for dense text retrieval.
\newblock In \emph{International Conference on Learning Representations}, 2021.

\bibitem[Yang et~al.(2018)Yang, Qi, Zhang, Bengio, Cohen, Salakhutdinov, and Manning]{yang2018hotpotqa}
Zhilin Yang, Peng Qi, Saizheng Zhang, Yoshua Bengio, William Cohen, Ruslan Salakhutdinov, and Christopher~D Manning.
\newblock Hotpotqa: A dataset for diverse, explainable multi-hop question answering.
\newblock In \emph{Proceedings of the 2018 Conference on Empirical Methods in Natural Language Processing}, 2018.

\bibitem[Yates et~al.(2021)Yates, Nogueira, and Lin]{yates2021pretrained}
Andrew Yates, Rodrigo Nogueira, and Jimmy Lin.
\newblock Pretrained transformers for text ranking: Bert and beyond.
\newblock In \emph{Proceedings of the 14th ACM International Conference on web search and data mining}, pp.\  1154--1156, 2021.

\bibitem[Yining et~al.(2013)Yining, Liwei, Yuanzhi, Di, Wei, and Tie-Yan]{yining2013theoretical}
Wang Yining, Wang Liwei, Li~Yuanzhi, He~Di, Chen Wei, and Liu Tie-Yan.
\newblock A theoretical analysis of ndcg ranking measures.
\newblock In \emph{Proceedings of the 26th annual conference on learning theory}, 2013.

\bibitem[Yu et~al.(2020)Yu, Twardowski, Liu, Herranz, Wang, Cheng, Jui, and Weijer]{Yu_2020_CVPR}
Lu~Yu, Bartlomiej Twardowski, Xialei Liu, Luis Herranz, Kai Wang, Yongmei Cheng, Shangling Jui, and Joost van~de Weijer.
\newblock Semantic drift compensation for class-incremental learning.
\newblock In \emph{Conference on Computer Vision and Pattern Recognition (CVPR)}, 2020.

\bibitem[Zhou et~al.(2024)Zhou, Wang, Qi, Ye, Zhan, and Liu]{zhou2024class}
Da-Wei Zhou, Qi-Wei Wang, Zhi-Hong Qi, Han-Jia Ye, De-Chuan Zhan, and Ziwei Liu.
\newblock Class-incremental learning: A survey.
\newblock \emph{IEEE Transactions on Pattern Analysis and Machine Intelligence}, 2024.

\end{thebibliography}
\bibliographystyle{collas2025_conference}

\newpage
\appendix

\section{Using QDC in Class-Incremental Learning Setting}
\label{app-cil}

It would be interesting to extend QDC to CIL setting with no access to task-id and future works could explore that setting. Similar to several existing works in CIL which predicts task-id (see discussion in~\cite{kim2025open}), the proposed method QDC could be adapted to CIL setting by predicting the task-id of a given query as discussed below.
\begin{itemize}
    \item After training on each task, the feature centroid of that task can be stored. Centroids of all old tasks could be updated by adding the drift vector $\Delta^{t-1 \rightarrow t}$ of the current task. So, at the end of task t, we have the task centroids for all tasks in the updated embedding space.
    \item During inference with queries extracted using the latest task model, we can predict the task-id based on cosine distance between query embedding and the task centroids.
    \item After predicting the task-id for queries, we can perform QDC to move the query back to the embedding space of the task.
\end{itemize}

\section{Performance evaluation with other metrics}
\label{app-metrics}
We show the performance of different methods using other metrics like recall@10 and MAP@10 in~\cref{table_1_recall,table_1_map}. We observe the same trend that using QDC outperforms FT and FT+KD and achieves similar performance as joint training.

\begin{table}[hbt!]
\begin{center}
\caption{Performance comparison of different approaches for Continual Document Retrieval tasks. Here, we report the \textbf{recall@10} scores for retrieval of each task. For continually trained methods, we use the latest model after training on T5 for retrieval of all tasks. We highlight the proposed QDC-based approaches in \colorbox{LightPurple}{purple}.}
\resizebox{\textwidth}{!}{
\begin{tabular}{c|c|c|c|c|c|c}
Method & T1 - MS MARCO & T2 - NQ & T3 - Hotpot QA & T4 - FEVER & T5 - FiQA2018 & Average  \\
\hline
\rule{0pt}{2ex}
Joint Training &  60.8 & 72.8 & 75.5 & 87.9 & 40.3 & 67.5 \\
\hline
\rule{0pt}{2ex} 
FT & 54.2 & 71.3 & 64.5 & 81.9 & 41.1 & 62.6 \\
\rowcolor{LightPurple} FT + QDC & 59.5 & 73.3 & 72.2 & 87.1 & 41.1 & 66.6 \\
\hline
\rule{0pt}{2ex} 
FT + KD & 58.2 & 73.3 & 70.6 & 78.2 & 41.4 & 64.3 \\
\rowcolor{LightPurple} FT + KD + QDC & 60.6 & 75.3 & 73.7 & 86.1 & 41.4 & \textbf{67.4} \\
\hline
\rule{0pt}{2ex} 
FT (with re-indexing) & 53.6 & 67.0 & 67.6 & 87.4 & 41.1 & 63.3 \\
FT + KD (with re-indexing) & 53.8 & 68.1 & 69.2 & 86.1 & 41.4 & 63.7\\

\hline
\end{tabular} 
}
\label{table_1_recall}
\end{center}
\end{table}

\begin{table}[hbt!]
\begin{center}
\caption{Performance comparison of different approaches for Continual Document Retrieval tasks. Here, we report the \textbf{MAP@10} scores for retrieval of each task. For continually trained methods, we use the latest model after training on T5 for retrieval of all tasks. We highlight the proposed QDC-based approaches in \colorbox{LightPurple}{purple}.}
\resizebox{\textwidth}{!}{
\begin{tabular}{c|c|c|c|c|c|c}
Method & T1 - MS MARCO & T2 - NQ & T3 - Hotpot QA & T4 - FEVER & T5 - FiQA2018 & Average  \\
\hline
\rule{0pt}{2ex}
Joint Training & 32.2 & 42.6 & 63.9 & 69.7 & 26.5 & 47.0 \\
\hline
\rule{0pt}{2ex} 
FT & 28.6 & 42.8 & 50.6 & 62.7 & 27.0 & 42.3 \\
\rowcolor{LightPurple} FT + QDC & 31.6 & 44.8 & 59.3 & 69.9 & 27.0 & 46.5 \\
\hline
\rule{0pt}{2ex} 
FT + KD & 30.9 & 44.9 & 57.4 & 58.1 & 26.9 & 43.6 \\
\rowcolor{LightPurple} FT + KD + QDC & 32.4 & 46.2 & 61.2 & 68.5 & 26.9 & \textbf{47.0} \\
\hline
\rule{0pt}{2ex} 
FT (with re-indexing) & 27.4 & 37.8 & 54.9 & 67.9 & 27.0 & 43.0 \\
FT + KD (with re-indexing) & 27.7 & 38.9 & 56.0 & 67.1 & 26.9 & 43.3\\

\hline
\end{tabular} 
}
\label{table_1_map}
\end{center}
\end{table}

\section{Comparison of Nomic pre-trained model with other retrieval models}
\label{app-nomic}
We compare the performance of the nomic retriever model trained either jointly or continually with classical dense retrievers like DPR~\citep{karpukhin2020dense}, ColBERT~\citep{khattab2020colbert} and ANCE~\citep{xiongapproximate}. We report performance of DPR, ColBERT and ANCE from BEIR~\citep{thakur2021beir}. While the other dense retriever models like ColBERT perform competitively, the nomic retriever model outperforms them. For the continual setting, we base the comparison on the nomic model. Benchmarking the performance of these other retriever architectures by continually training them in the proposed continual setting could be interesting to explore in future works.

\begin{table}
\begin{center}
\caption{Performance comparison of different approaches. We report the nDCG@10 scores for retrieval of each task. Here, we do not consider continual training and evaluate the performance on each task separately. $\dagger$: results excerpted from~\cite{thakur2021beir}.}
\resizebox{0.95\textwidth}{!}{
\begin{tabular}{c|c|c|c|c|c|c|c}

Method & Continual & T1 - MS MARCO & T2 - NQ & T3 - Hotpot QA & T4 - FEVER & T5 - FiQA2018 & Avg. Score  \\
\hline
\rule{0pt}{2ex}
BM25$^\dagger$ & \XSolidBrush & 22.8 & 32.9 & 60.3 & 75.3 & 23.6 & 43.0 \\
DPR$^\dagger$ & \XSolidBrush & 17.7 & 47.4 & 39.1 & 56.2 & 11.2 & 34.3 \\
ANCE$^\dagger$ & \XSolidBrush & 38.8 & 44.6 & 45.6 & 66.9 & 29.5 & 45.1 \\
ColBERT$^\dagger$ & \XSolidBrush & 40.1 & 52.4 & 59.3 & 77.1 & 31.7 & 52.1 \\
\midrule
Joint Training (Nomic) & \XSolidBrush & 39.2 & 50.7 & 71.7 & 75.1 & 33.8 & \textbf{54.1} \\
FT + KD + QDC (Nomic) & \Checkmark & 39.3 & 54.1 & 69.3 & 73.6 & 34.3 & \textbf{54.1} \\
\hline
\end{tabular} 
}
\label{table_nomic}
\end{center}
\vspace{-3mm} 
\end{table}

\section{Retrieval examples} 
\label{app-examples}
We present some retrieval results using the continually fine-tuned model in~\cref{table:examples_marco,table:examples_nq} for old tasks (MS MARCO and Hotpot QA). We show that using the proposed QDC method with fine-tuned model retrieves more relevant documents from the corpus for a given query.

\begin{table}[hbt!]
\begin{center}
\caption{Examples of top-1 retrieved document for a given query using fine-tuned model and with the proposed QDC (FT+QDC) for \textbf{MS MARCO} dataset using the  continually trained model after task 5.}
\resizebox{\textwidth}{!}{
\begin{tabular}{c|c|c}
Query & Document retrieved with FT & Document retrieved with FT+QDC \\
\hline
\rule{0pt}{2ex} 
do vhi swiftcare do  & Blood tests. Information on having blood tests and the types of blood &  The SwiftCare clinics charge an initial consultation fee of 85 euro, with additional \\
blood tests? & tests you might have. Your blood sample is sent to the laboratory. &  charges for tests and procedures. For example, an x-ray at the clinics costs 65 euro, \\
 & A blood doctor can look at your sample under a microscope. They can &  blood tests range from 30 to 50 euro and complex suturing costs 50 euro.  \\
 & see the different types of cells and can count the different blood cells. & The Swiftcare clinics are staffed by doctors with significant experience in general \\
 & & practice and emergency care, according to VHI. The clinics are run as a joint \\
 & & initiative between the VHI and The Well, a primary care care medical company. \\
\hline
the miners state bank & Search all THE MINERS STATE BANK routing numbers in the table below. & The Miners State Bank Routing Number the miners state bank routing aba number  \\
routing number & Use the Search box to filter by city, state, address, routing number. & 091109253 routing number is a 9-digit number designed and assigned to    \\
 & Click on the routing number link in the table below to navigate to it and & The Miners State Bank by The American Bankers Association (ABA) to \\
 & see all the information about it (address, telephone number, zip code, etc.). & identify the financial institution upon which a payment was drawn.  \\
\hline
cadillac alternator price & 1 On average, a car alternator prices are going to range anywhere from $66 to $320. & A Cadillac De Ville Alternator Replacement costs between $378 and $860 on average.  \\
& 2  This is not going to include the labor costs. 3  When you factor in labor costs, & Get a free detailed estimate for a repair in your area. A Cadillac De Ville \\
& itâ safe to add another $100 to $275. & Alternator Replacement costs between $378 and $860 on average. \\
\hline
door knocker definition & Definition of 'knocker'. knocker. A knocker is a piece of metal on the front  & Door Knocker definition. An act of physical violence performed on a person (usually  \\
& door of a building, which you use to hit the door in order to  &  a woman) who is wearing huge hoop earrings. A cob of dried corn employed by  \\
& attract the attention of the people inside. &  pranksters on Mischief evening to toss at people's doors......notable for  \\
& & loud BLANG!! or KABAAM!! Not what title really seems. \\
\hline
\end{tabular} 
}
\label{table:examples_marco}
\end{center}
\end{table}

\begin{table}
\begin{center}
\caption{Examples of top-1 retrieved document for a given query using fine-tuned model and with the proposed QDC (FT+QDC) for \textbf{Hotpot QA} dataset using the continually trained model after task 5.}
\resizebox{\textwidth}{!}{
\begin{tabular}{c|c|c}
Query & Document retrieved with FT & Document retrieved with FT+QDC \\
\hline
\rule{0pt}{2ex} 
This singer of A Rather & Pinocchio (singer) Pinocchio is a fictional, animated French character & A Rather Blustery Day \"A Rather Blustery Day\" is a whimsical song \\
Blustery Day also voiced &  and singer. & from the Walt Disney musical film featurette, \"Winnie the Pooh \\
what hedgehog? &  & and the Blustery Day\". It was written by Robert  \& Richard Sherman \\
&  & and sung by Jim Cummings as \"Pooh\". \\
\hline
What WB supernatrual drama & Charmed (season 4) The fourth season of \"Charmed\", an American  & Rose McGowan Rose Arianna McGowan (born September 5, 1973)  \\
series was Jawbreaker star & supernatural drama television series, began airing on October 4, 2001  & is an Italian-born American actress, film producer, director and singer.  \\
Rose Mcgowan best known & on The WB. Airing on Thursdays at 9:00 pm, the season consisted of 22  & She is best known to television audiences for having played Paige   \\
for being in? & episodes and concluded its airing on May 16, 2002. This season also saw  & Matthews in The WB supernatural drama series \"Charmed\" from 2001  \\
& the introduction of Rose McGowan as Paige Matthews—half-sister to Prue, & to 2006. \\
& Piper and Phoebe—and a slight alteration of the opening credits, due to &  \\
& the third season departure of Shannen Doherty as Prue. Paramount Home &  \\
& Entertainment released the complete fourth season in a six-disc &  \\
& boxed set on February 28, 2006. &  \\
\hline
In which role did Caroline & The Magical Legend of the Leprechauns The Magical Legend of the  & Caroline Carver (actress) Caroline Carver (born 1976) is an English \\
Carver played in a 1999 Hallmark & Leprechauns is a 1999 Hallmark Entertainment made-for-TV fantasy movie.  & actress, screenwriter, and producer best known for roles such as  \\
Entertainment made-for-TV & It stars Randy Quaid, Colm Meaney, Kieran Culkin, Roger Daltrey, Caroline  & Princess Jessica in the TV film \"The Magical Legend of the  \\
fantasy movie? & and Whoopi Goldberg. The film contains two main stories that eventually  & Leprechauns\" (1999), Ingrid in \"The Aryan Couple\" (2004), and Sandy \\
&  Carver intertwine: the first being the story of an American businessman who  &  in \"My First Wedding\" (2006).  \\
&  visits Ireland and encounters magical leprechauns and the second, a story of a  &  \\
& pair of star-crossed lovers who happen to be a fairy and a leprechaun, &  \\
& belonging to opposing sides of a magical war. It contains many references to  &  \\
& Romeo and Juliet such as two lovers taking poison and feuding clans. &  \\
\hline
What's the name of the fantasy film & Fredegar Bolger Fredegar \"Fatty\" Bolger is a fictional character  & Sarah Bolger Sarah Lee Bolger (born 28 February 1991) is an Irish  \\
starring Sarah Bolger, featuring a & in J. R. R. Tolkien's fantasy novel \"The Lord of the Rings\". & actress. She is best known for her roles in the films \"In America\",  \\
New England family who discover &  & \"Stormbreaker\", and \"The Spiderwick Chronicles\", as well as her  \\
magical creatures around their estate? &  & award-winning role as Lady Mary Tudor in the TV series \"The Tudors\", \\
&  &  and for guest starring as Princess Aurora in \"Once Upon a Time\". \\
\hline
\end{tabular} 
}
\label{table:examples_nq}
\end{center}
\end{table}

\end{document}